\documentclass[preprint2]{aastex}

\usepackage{graphics,latexsym,psfig}

\def\etal{{et\,al.}}
\def\msun{M$_{\odot}$}

\def\mdot{$\dot M$}
\def\grad{$^\circ$}
\def\degs{\ifmmode ^{\circ}\else$^{\circ}$\fi}
\def\amin{\ifmmode ^{\prime}\else$^{\prime}$\fi}
\def\asec{\ifmmode ^{\prime\prime}\else$^{\prime\prime}$\fi}
\def\fss{\hbox{$.\!\!^{\rm s}$}}        
\def\fdg{\hbox{$.\!\!^\circ$}}          
\def\farcs{\hbox{$.\!\!^{\prime\prime}$}}  
\newbox\grsign \setbox\grsign=\hbox{$>$}
\newdimen\grdimen \grdimen=\ht\grsign
\newbox\laxbox \newbox\gaxbox
\setbox\gaxbox=\hbox{\raise.5ex\hbox{$>$}\llap
     {\lower.5ex\hbox{$\sim$}}}\ht1=\grdimen\dp1=0pt
\setbox\laxbox=\hbox{\raise.5ex\hbox{$<$}\llap
     {\lower.5ex\hbox{$\sim$}}}\ht2=\grdimen\dp2=0pt

\def\lax{$\mathrel{\copy\laxbox}$}
\def\gr{\hbox{ \raisebox{-1.0mm}{$\stackrel{>}{\sim}$} }}

\def\h{$^{\rm h}$}\def\m{$^{\rm m}$}

\begin{document}

\title{GRB 011121: A collimated outflow into wind-blown
       surroundings\thanks{Based on observations collected   at  the
       European Southern   Observatory,   La Silla and Paranal, Chile  (ESO
       Programme  165.H-0464).}}



\author{J. Greiner$^{1,2}$,  S. Klose$^{3}$,  M. Salvato$^{1,2}$, 
    A. Zeh$^{3}$, R. Schwarz$^{1,4}$,
      D.H. Hartmann$^{5}$,  N. Masetti$^{6}$,
     B. Stecklum$^{3}$,
     G. Lamer$^{1}$, N. Lodieu$^{1}$,  R.D. Scholz$^{1}$,
     C. Sterken$^{7}$, J. Gorosabel$^{8, 9}$,
     I. Burud$^{9}$,
     J. Rhoads$^{9}$, 
     I. Mitrofanov$^{10}$, M. Litvak$^{10}$,
           A. Sanin$^{10}$, V. Grinkov$^{10}$,
     M.I. Andersen$^{1}$,
     J.M. Castro Cer\'on$^{11}$,
     A.J. Castro-Tirado$^{8,12}$,
     A. Fruchter$^{9}$,
     J.U. Fynbo$^{13}$,
     J. Hjorth$^{14}$,
     L. Kaper$^{15}$,
     C. Kouveliotou$^{16}$, 
     E. Palazzi$^{6}$,
     E. Pian$^{17}$,
     E. Rol$^{15}$,
     N.R. Tanvir$^{18}$,
     P.M. Vreeswijk$^{19}$,
     R.A.M.J. Wijers$^{15}$,
     E. van den Heuvel$^{15}$ \\}


 \affil{$^{1}$ Astrophysikalisches Institut, 14482 Potsdam, Germany \\
   $^{2}$ Max-Planck-Institut f\"ur extraterrestrische Physik, 85741
   Garching, Germany \\ 
   $^{3}$ Th\"uringer Landessternwarte Tautenburg,
   07778 Tautenburg, Germany \\
  $^{4}$ Universit\"ats-Sternwarte
   G\"ottingen, Geismarlandstr. 11,  37083 G\"ottingen, Germany \\
   $^{5}$ Clemson University, Department of Physics and Astronomy, Clemson,
   SC 29634, USA \\
   $^{6}$ Istituto di Astrofisica Spaziale e
   Fisica Cosmica, CNR, Sez. di Bologna, Via Gobetti 101, 40129 Bologna,
    Italy \\
   $^{7}$  University of Brussels (VUB), Pleinlaan 2, 1050
   Brussels, Belgium \\ 
   $^{8}$ Instituto de Astrof\'{\i}sica de Andaluc\'{\i}a (IAA-CSIC),
   P.O. Box 03004, 18080 Granada, Spain \\
  $^{9}$ Space Telescope Science Institute, 3700 San
   Martin Drive, Baltimore,  MD 21218, USA \\
  $^{10}$ Space Research
   Institute, Profsoyusnaya 84/32, 117810 Moscow, Russia \\
    $^{11}$ Real Instituto y Observatorio de la Armada, Secci\'on
   de Astronom\'{\i}a, 11.110 San Fernando-Naval (C\'adiz), Spain \\
    $^{12}$ Laboratorio de Astrof\'{\i}sica Espacial y
   F\'{\i}sica Fundamental,  
   Madrid, Spain \\
    $^{13}$ Department of Physics and Astronomy, University of Aarhus, 
       Ny Munkegade, 8000 Aarhus C, Denmark \\
    $^{14}$ Astronomical Observatory, University of Copenhagen, 
      Juliane Maries Vej 30, 2100 Copenhagen, Denmark \\ 
    $^{15}$ University of Amsterdam,
     Kruislaan 403, 1098 SJ Amsterdam, The Netherlands \\
    $^{16}$ NSSTC, SD-50, 320 Sparkman Drive,
     Huntsville, AL 35805, U.S.A. \\
     $^{17}$ INAF, Osservatorio Astronomico di Trieste, Via
     Tiepolo 11, 34131 Trieste, Italy \\
    $^{18}$ Department of Physical Sciences, Univ. of Hertfordshire, 
     College Lane, Hatfield Herts, AL10 9AB, UK \\
     $^{19}$ European Southern Observatory, Alonso de Cordova 3107, Vitacura,
     Casilla 19001, Santiago 19, Chile \\}


\begin{abstract}
We report optical and near-infrared follow-up observations of GRB
 011121 collected predominantly at ESO telescopes in Chile. We discover a
 break in the afterglow light curve after 1.3 days, which implies an initial
 jet opening angle of about 9\degr. The jet origin of this break is supported
 by the fact that the spectral energy distribution is achromatic during the
 first four days.  During later phases, GRB 011121 shows significant excess
 emission above the flux predicted by a power law, which we interpret as
 additional light from an underlying supernova. 
In particular, the spectral energy distribution of the
optical transient  approximately 2 weeks after the burst is clearly
not of power-law type, but can be presented by a black body with a
temperature of $\sim$6000 K.
The deduced parameters for the
 decay slope as well as the spectral index favor  a wind scenario, i.e. an
 outflow into a circum-burst environment shaped by the stellar wind of a
 massive GRB progenitor. Due to its low redshift of  $z$=0.36, GRB 011121 
has been the best example for the GRB-supernova connection
 until GRB 030329,  and provides
 compelling evidence for a circum-burster wind region expected to exist if the
 progenitor was a massive star.
\end{abstract}
\keywords{gamma rays: bursts --- techniques: photometric --- supernovae: general}

%
%


\section{Introduction}

In the presently favoured scenario for classical gamma-ray bursts (GRBs),
occurring at cosmological distances (measured redshift range so far 
$0.36 < z < 4.5$; van Paradijs \etal\ 1997; Andersen \etal\ 2000),
the explosion of a very massive star leads to a fireball and a
short, beamed flash of gamma-rays (Woosley 1993; Fryer \etal\ 1999;
M\'esz\'aros 2002),
resulting in three physically distinct observational phenomena,
namely the GRB itself, a long-lasting afterglow and the classical
supernova (SN) light. Whereas the afterglow emission is probably fed by 
the kinetic energy of the collimated, relativistic outflow, the supernova
light is caused by the decay of radioactive nuclei created and released
during the stellar explosion. The maximum of the supernova light is 
expected at $\sim$10--20 (1+$z$) days after the explosion, though at
present it is not clear whether or not the GRB and the supernova explosion 
are delayed (Vietri \& Stella 1998). One of the basic consequences of this
hypernova scenario is that it predicts (and links) the occurrence
of a GRB with a jet, the unavoidable strong wind from the massive 
progenitor star, and the supernova light (Heger \etal\ 2003).


\begin{figure}[ht]
 \vbox{\psfig{figure=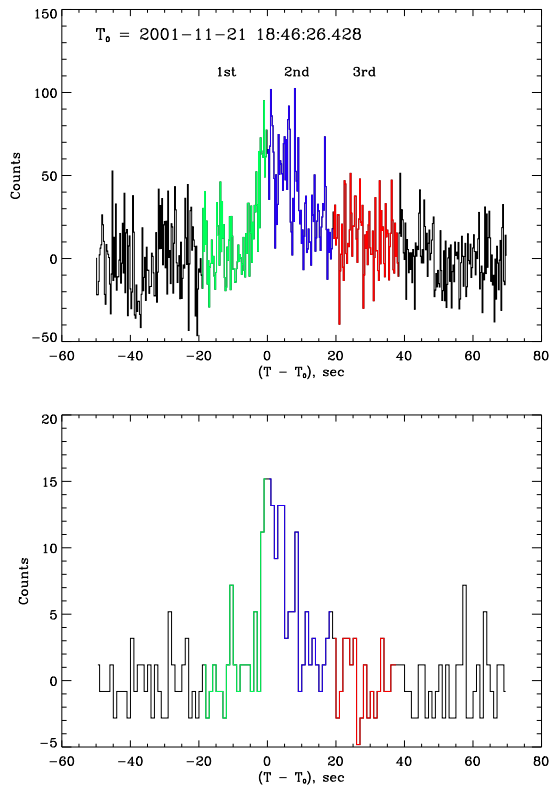,width=7.7cm}}
    \caption[hendlc]{Light curve of GRB 011121 as recorded by HEND on
      Mars Odyssey in the outer scintillator at $>$30 keV with 0.25 sec
      time resolution (top) and in the inner scintillator at $>$60 keV 
      with 1 sec time resolution (bottom). The time axis 
      zero point (UT) is indicated at the top. The three colours represent the
      three successive time segments used for the spectral fitting 
      (see Fig. \ref{xspec}).
    \label{hend}}
    \end{figure}


Earlier evidence on the GRB-SN connection was initially based on the
coincidence of GRB 980425 and SN1998bw (Galama \etal\ 1998), and subsequently
some supernova light contribution was found in the late-time light curves of
GRB 980326 (Castro-Tirado \& Gorosabel 1999, Bloom \etal\ 1999), GRB 970228
(Galama \etal\ 2000), GRB 970508 (Sokolov \etal\ 2002)  and possibly GRB
980703 (Holland \etal\ 2001), GRB 990712 (Bj\"ornsson \etal\ 2001),  GRB
991208 (Castro-Tirado \etal\ 2001) and GRB 000911 (Lazzati \etal\ 2001).

Observations of massive stars in our Galaxy have shown that they lose
matter via strong stellar winds in the LBV and in the Wolf-Rayet
evolutionary phase. GRB progenitors are expected to be very massive
stars  and perhaps physically related to Ib/c supernovae (Heger \etal\ 2003).
The progenitors of these stars should develop fast winds at the end of
their lives and thus, wind features are expected to be seen in   
afterglow light curves. The evolution of classical supernova remnants
into wind-blown bubbles has been a target of detailed investigations
since years (e.g., Benetti et al. 1999; Landecker et al. 1999).  What
has been missing in GRB research until GRB 030329 (Hjorth \etal\ 2003),
however, was a strong
observational link between an underlying supernova component in an
afterglow light curve and evidence for a fireball expanding into a 
stellar wind profile.  Significant work into this direction has been
done over the past years by many groups from the theoretical as well
as from the observational side (e.g., M\'esz\'aros, Rees, \& Wijers
1998; Chevalier \& Li 2000) and GRBs with wind-profile interactions have 
been identified
in some cases, including GRB 980425/SN 1998bw (Chevalier \& Li 2000).
However, if one excludes GRB 980425 from this list, prior to GRB
011121 the observational data base was rather poor. The burst 011121
changed this observational situation, though in a less spectacular way
than GRB 030329.
Not only was
the  distance of the burster relatively small but also the light curve
steepened 1 day after the burst so that a bright supernova component 
became visible.

GRB 011121 was detected by the GRBM/WFC onboard BeppoSAX on 2001
November 21, 18:47:21 UT and localized to initially 5 arcmin 
(Piro 2001a). Subsequent analysis refined this position to 2 arcmin
accuracy (Piro 2001b), and the triangulation of the GRB arrival
times as measured by Ulysses, BeppoSAX  (Hurley \etal\ 2001) and
HEND (Hurley \etal\ 2002) further refined the coordinates. A follow-up
X-ray observation with the BeppoSAX narrow-field instruments revealed
a fading X-ray afterglow (Piro \etal\ 2001).

Follow-up optical/NIR observations of GRB 011121 were quickly started
by several groups, leading to independent discoveries of the afterglow
(e.g., Wyrzykowski \etal\ 2001; Greiner \etal\ 2001). Further observations
revealed a rather small distance, $z \approx 0.36$ (Infante \etal\ 2001),
and excess emission above the early power-law decay (Garnavich \etal\ 2001).
The interpretation of this excess light as a possible supernova bump 
attracted much attention and resulted in the so far best sampling of 
GRB afterglow emission at late times.

Here we report the results of observations obtained by the 
GRACE\footnote{\tt http://zon.wins.uva.nl/$\sim$grb/grace/}
(GRB Afterglow Collaboration at ESO) consortium.

\section{Observations}

\subsection{X-rays/Gamma-rays \label{gamma}}

The light curves of GRB 011121 measured with the two detectors of HEND are
shown in Fig. \ref{hend}.  Integrated into the Third Interplanetary Network,
the ``Mars Odyssey'' satellite  has two instruments with GRB detection
capabilities: the Gamma-Ray Spectrometer and the High Energy Neutron Detector
(HEND).  HEND combines four detectors to measure the spectra of neutrons and
gamma-rays (Mitrofanov \etal\ 2003).  In particular, the inner/outer
CsI-scintillators can measure gamma-ray  photons in the 30--1000 keV range, at
commandable integration times, and in case of a GRB-trigger (or solar flare),
time histories at 1.0/0.25 sec resolution are recorded.

Fig. \ref{hend} shows that the gamma-ray emission is characterized by a
main peak with $\sim$10 sec duration, followed by a $\sim$20 sec tail.
Thus, GRB 011121 clearly belongs to the long-duration sub-class of GRBs.


 \begin{figure}[ht]
  \vbox{\psfig{figure=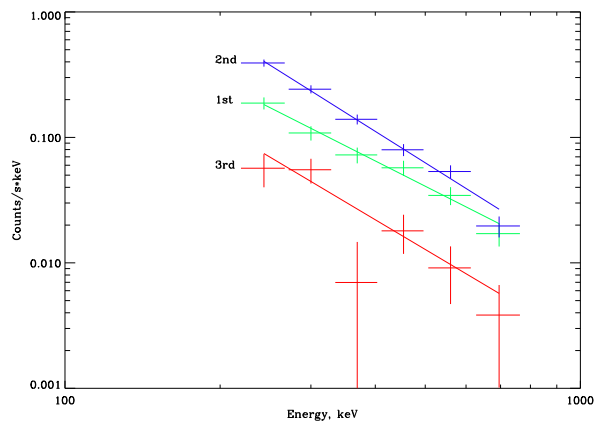,width=8.cm,angle=90}}
    \caption[hendspec]{Energy spectra of GRB 011121 above 200 keV,
       as recorded by HEND on Mars Odyssey in the outer scintillator 
       for three successive time segments, shown as green, blue and red 
        in the light curve (see Fig. \ref{hend}.)
    \label{xspec}}
    \end{figure}


According to the HEND energy spectra (Fig. \ref{xspec}) there is no
obvious spectral evolution along the light curve of GRB 011121.
All three 20 sec segments of the time profile have spectra with the same 
power law slope with a photon index of --2.35$\pm$0.25 (after backwards
folding of the model with the instrument response). 
There is no evidence
in the HEND data that the early afterglow emission started already at the
time of the last segment of the burst.

Based on the power law spectral slope of 2.35,
the fluence is $2\pm0.4 \times 10^{-5}$ erg cm$^{-2}$ in the
250--700 keV range. 
This corresponds to an (isotropic) energy release of 
$E$ = 2.7$\times$10$^{52}$ erg at the given redshift (luminosity distance
of 2.07 Gpc) and 
the cosmological parameters  $\Omega_\Lambda = 0.7, \Omega_M$ =
0.3, and $H_0 = 65$ km s$^{-1}$ Mpc$^{-1}$.


\begin{figure}[h]
 \vbox{\psfig{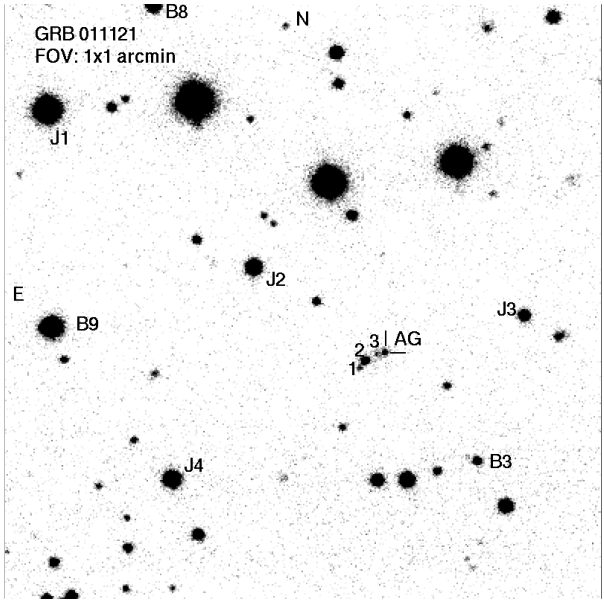}}\par
\caption[jfc]{$J$ band finding chart of GRB 011121 obtained on 2001 Nov. 24
    with ISAAC/VLT. Marked are the afterglow (AG), neighbouring objects 
    (1, 2), the host galaxy (3) and some local standards (J1--J4, B3, B8, B9).
    \label{jfc}}
\end{figure}


\subsection{Optical and NIR Imaging}

About 9 hours after the GRB, $K_s$- and $R$-band images were obtained 
at La Silla (ESO) with SOFI
(equipped with a 1024$\,\times\,$1024 HgCdTe HAWAII  array with 18.5
$\mu$m pixels and a plate scale of  0\farcs29/pixel) at the 3.58m NTT, and
with  EFOSC2 (equipped with a 2048$\,\times\,$2048 thinned Loral/Lesser
CCD with  15 $\mu$m pixels and a plate scale of 0\farcs157/pixel) at
the 3.6m telescope, respectively. After our independent discovery of
the afterglow, second epoch observations plus some exposures with 
additional filters were obtained ($\sim$ 4 hrs later) at the end of that
night. Multi-colour photometry continued at the NTT and 3.6\,m telescope 
during the next three days and at the VLT thereafter. 
Due to the rapidly decreasing afterglow brightness we switched to
the instruments at the 8.2\,m Very Large Telescopes (VLT) starting on
day 3 after the GRB. For our near-infrared (NIR)
observations, the short wavelength (0.9--2.5 $\mu$m) arm of the
infrared spectrometer ISAAC (equipped with a 1024$\times$1024 pixel
Rockwell HgCdTe array and a plate scale of 0\farcs147/pixel) on the
VLT-telescope Antu was used, and for the optical observations the
Focal Reducer and Spectrograph FORS (equipped with a SITE CCD with 24
$\mu$m pixels and a plate scale of 0\farcs2/pixel) at VLT/Yepun.


Additional imaging was performed in the $R$ and $I$ bands at the 
ESO 1.54\,m Danish telescope on La Silla (Chile), in the $H$ band at the 
Anglo-Australian Telescope (AAT) and in the $J$ and $H$ bands at 
the 4-m Blanco telescope of the Cerro Tololo Interamerican Observatory (CTIO).
The 1.54\,m Danish telescope was equipped with a 2048$\times$4096 EEV/MAT
frame-transfer CCD (illuminated area is  2048$\times$2048 pixel)
with  15 $\mu$m pixels, which provides a plate scale of 0\farcs39/pixel.
At the AAT the IRIS2 instrument with a 1024$\times$1024 Rockwell HAWAII-1 
HgCdTe detector was used, 
having a pixel scale of 0\farcs446/pixel. At CTIO,  the Ohio State 
InfraRed Imager/Spectrometer (OSIRIS) 
with a 1024$\times$1024 HAWAII-1 HgCdTe detector was
used, having a plate scale of 0\farcs161/pixel.


\begin{table*}
\caption{Log of the observations \label{log}}
\footnotesize
\tiny{
\begin{tabular}{llcccc}
   \hline
   \noalign{\smallskip}
   ~~~~~~~Date (UT) & Telescope/Instrument & Filter/Grism$^{(a)}$
  & Exposure (sec) &   Seeing & Brightness (mag)$^{(b)}$ \\
   \noalign{\smallskip}
   \hline
   \noalign{\smallskip}
 2001 Nov 22 03:03--03:25 & ESO NTT/SOFI     & $K_{\rm s}$ &  10$\times$63 & 0\farcs7
   & 15.19$\pm$0.07 \\
 2001 Nov 22 04:03--04:14 & ESO 3.6\,m/EFOSC2 & $R_{\rm c}$  &  3$\times$180 & 1\farcs5
   & 18.86$\pm$0.02  \\
 2001 Nov 22 07:30--08:01 & ESO 3.6\,m/EFOSC2 & $R_{\rm c}$  &  3$\times$600 & 1\farcs0
   & 19.51$\pm$0.01 \\
 2001 Nov 22 07:47--08:22 & ESO NTT/SOFI     & $J$ &  25$\times$70 & 0\farcs8
    &  17.47$\pm$0.04 \\
 2001 Nov 22 08:23--08:47 & ESO NTT/SOFI     & $H$ &  17$\times$70 & 0\farcs7
    & 16.72$\pm$0.05 \\
 2001 Nov 22 08:47--09:06 & ESO NTT/SOFI     & $K_{\rm s}$ &  13$\times$70 & 0\farcs9
    &  16.09$\pm$0.07 \\
 2001 Nov 22 14:00--14:14 & AAO AAT/IRIS2 & $H$ & 10$\times$60 & 2\farcs7
    & 17.30$\pm$0.25 \\
 2001 Nov 23 00:04--00:20 & ESO NTT/SOFI  & $K_{\rm s}$ & 11$\times$70 & 1\farcs2
    & 17.71$\pm$0.10 \\
 2001 Nov 23 00:20--00:39 & ESO NTT/SOFI  & $J$  & 13$\times$70 & 1\farcs6
   &  19.38$\pm$0.10 \\
 2001 Nov 23 04:30--04:40 & ESO 3.6\,m/EFOSC2 & $R_{\rm c}$ & 3$\times$180 & $^{(c)}$
   & 21.23$\pm$0.05 \\
 2001 Nov 23 04:41--04:57 & ESO 3.6\,m/EFOSC2 & $B$ & 3$\times$300 &  $^{(c)}$
   &  22.40$\pm$0.50 \\
 2001 Nov 23 04:58--05:08 & ESO 3.6\,m/EFOSC2 & $V$ & 3$\times$180 &  $^{(c)}$
   & 21.95$\pm$0.15  \\
 2001 Nov 23 08:29--08:45 & ESO NTT/SOFI  & $J$ & 11$\times$70 & 0\farcs9
    & 19.73$\pm$0.10 \\
 2001 Nov 23 08:45--09:01 & ESO NTT/SOFI  & $K_{\rm s}$ & 11$\times$70 & 0\farcs9
    & 18.29$\pm$0.10 \\
 2001 Nov 23 08:29--08:39 & ESO 3.6\,m/EFOSC2 & $R_{\rm c}$ & 3$\times$180 & 1\farcs3
    & 21.75$\pm$0.08 \\
 2001 Nov 23 08:39--08:44 & ESO 3.6\,m/EFOSC2 & $B$ & 300 & 1\farcs6
    & $>$21.70 \\
 2001 Nov 23 06:45--08:34 & ESO Yepun/FORS2 & 600B & 2$\times$2400+1500 &
  1\farcs5,1\farcs0,1\farcs2  & ---\\
 2001 Nov 23 07:22--08:34 & ESO Antu/ISAAC & 1.1--1.4 $\mu$m &  3600 & 1\farcs2& --- \\
 2001 Nov 24 06:24--06:48 & ESO Antu/ISAAC & $K_{\rm s}$ & 15$\times$60 & 0\farcs7
   & 19.30$\pm$0.08 \\
 2001 Nov 24 06:53--07:16 & ESO Antu/ISAAC & $H$ & 15$\times$60 & 0\farcs7
    & 19.81$\pm$0.05 \\
 2001 Nov 24 07:18--07:36 & ESO Antu/ISAAC & $J_{\rm s}$ &  10$\times$90 & 0\farcs6
   & 20.93$\pm$0.10 \\
 2001 Nov 24 07:53--08:59 & ESO Antu/ISAAC & 1.95--2.55 $\mu$m & 3600 & 1\farcs2& ---\\
 2001 Nov 24 07:43--08:01 & ESO Yepun/FORS2 & $B$  & 3$\times$300 & 0\farcs8
    & 23.70$\pm$0.30 \\
 2001 Nov 24 08:02--08:14 & ESO Yepun/FORS2 & $V$  & 3$\times$180 & 0\farcs8 &
  23.37$\pm$0.05  \\
 2001 Nov 24 08:15--08:27 & ESO Yepun/FORS2 & $R_{\rm special}$ & 3$\times$180 & 0\farcs8 & 22.93$\pm$0.08 \\
 2001 Nov 25 06:43--06:59 & ESO Antu/ISAAC & $J_{\rm s}$ &  10$\times$90 & 0\farcs9
    & 21.64$\pm$0.30  \\
 2001 Nov 25 07:01--07:24 & ESO Antu/ISAAC & $H$ &  15$\times$60 & 0\farcs6
    & 20.73$\pm$0.10  \\
 2001 Nov 25 07:27--07:52 & ESO Antu/ISAAC & $K_{\rm s}$ &  15$\times$60 & 0\farcs8
    & 20.02$\pm$0.10  \\
 2001 Nov 25 06:34--08:10 & ESO Yepun/FORS2 & 150I & 3$\times$1800 & 0\farcs5  & --- \\
 2001 Nov 25 08:10--08:22 & ESO Yepun/FORS2 & $R_{\rm special}$ & 3$\times$180 & 0\farcs8 & 23.68$\pm$0.15 \\
 2001 Nov 25 08:23--08:34 & ESO Yepun/FORS2 & $V$ & 3$\times$180 &  $^{(d)}$ &
 $>$24.40 \\
 2001 Nov 25 07:41--08:13 & ESO 1.54\,m Danish & $R_c$ & 1880 & 1\farcs1  & 23.60$\pm$0.40 \\
 2001 Nov 30 08:05--08:51 & CTIO OSIRIS     & $H$    & 19$\times$60 & 0\farcs7  & $>$19.40 \\
 2001 Dec 01 08:07--08:39 & CTIO OSIRIS     & $J$    & 8$\times$120 & 0\farcs9  & $>$20.10 \\
 2001 Dec 03 07:50--08:25 & CTIO OSIRIS     & $J$    & 10$\times$120 & 1\farcs2 & $>$20.10 \\
 2001 Dec 05 06:28--07:01 & ESO Melipal/FORS1 & $V$  & 3$\times$600 & 0\farcs9  & 24.30$\pm$0.02 \\
 2001 Dec 05 07:02--07:33 & ESO Melipal/FORS1 & $R_{\rm c}$  & 3$\times$600 & 0\farcs9&
  23.20$\pm$0.08 \\
 2001 Dec 05 07:34--08:07 & ESO Melipal/FORS1 & $I_{\rm c}$  & 3$\times$600 & 0\farcs9&
 22.35$\pm$0.25  \\
 2001 Dec 05 06:53--08:09 & ESO Antu/ISAAC  & $J_{\rm s}$ & 40$\times$90 & 0\farcs6
   & 22.41$\pm$0.15 \\
 2001 Dec 07 06:45--08:02 & ESO Antu/ISAAC  & $J_{\rm s}$ & 40$\times$90 & 0\farcs6
   & 22.69$\pm$0.10  \\
 2001 Dec 08 06:20--06:30 & ESO Melipal/FORS1 & $V$  & 600 & 0\farcs8  &
   24.09$\pm$0.02 \\
 2001 Dec 09 06:49--08:05 & ESO Antu/ISAAC  & $J_{\rm s}$ & 40$\times$90 & 0\farcs7
   & 22.52$\pm$0.15 \\
 2001 Dec 09 07:33--07:43 & ESO Melipal/FORS1 & $V$  & 600 & 1\farcs0  &
  23.94$\pm$0.02 \\
 2001 Dec 10 06:47--06:57 & ESO Melipal/FORS1 & $R_{\rm c}$  & 600 &  0\farcs8 &
  23.65$\pm$0.08\\
 2001 Dec 11 07:33--08:06 & ESO Melipal/FORS1 & $R_{\rm c}$  & 3$\times$600 & 0\farcs6
   & 23.45$\pm$0.08 \\
 2001 Dec 11 08:07--08:29 & ESO Melipal/FORS1 & $V$  & 2$\times$600 & 0\farcs7 & 24.18$\pm$0.02 \\
 2001 Dec 12 05:12--07:30 & ESO Yepun/FORS2 & 300I & 4$\times$1800 & 0\farcs7-1\farcs2 & ---\\
 2001 Dec 13 07:19--08:46 & ESO Antu/ISAAC  & $J_{\rm s}$ & 40$\times$90 & 0\farcs7
    & 22.89$\pm$0.15 \\
 2001 Dec 15 05:10--05:27 & ESO 1.54m Danish & $I_{\rm c}$ & 4$\times$900 & 1\farcs2
    & $>$24.70 \\
 2001 Dec 16 04:51--05:08 & ESO 1.54m Danish & $I_{\rm c}$ & 2$\times$900 & 1\farcs3
    & $>$23.90 \\
 2001 Dec 17 04:55--05:12 & ESO 1.54m Danish & $I_{\rm c}$ & 4$\times$900 & 2\farcs0
    & $>$23.80 \\
 2001 Dec 17 07:20--08:47 & ESO Antu/ISAAC  & $J_{\rm s}$ & 40$\times$90 & 1\farcs0
    & 22.79$\pm$0.30 \\
 2001 Dec 18 04:45--05:01 & ESO 1.54m Danish & $I_{\rm c}$ & 2$\times$900 & 1\farcs1
    & $>$23.80 \\
 2002 Feb 09 04:04--05:24 & ESO Antu/ISAAC  & $J_{\rm s}$ & 40$\times$90 & 0\farcs5
    & $>$24.85 \\
   \noalign{\smallskip}
   \hline
\end{tabular}
}

\noindent{$^{(a)}$ The $R_{\rm special}$ filter is about
10\% broader than the standard $R_{\rm c}$. Filters $J_{\rm s}$ and $K_{\rm s}$
are narrower with respect to the canonical $J$ and $K$ bands,
respectively: $J_{\rm s}$ has a width of 0.16 $\mu$m (instead of 0.29
$\mu$m), and $K_{\rm s}$ has a  width of  0.29 $\mu$m centered at 2.16
$\mu$m  (instead of 0.35 $\mu$m centered at 2.20 $\mu$m).
However, since both narrow-band filters have a higher transmission than the
canonical filters, the net effect is that 
$J-J_{\rm s}$ \lax\ 0.05 ($K-K_{\rm s}$ \lax\ 0.02)
when comparing  the ISAAC $J_{\rm s}$ vs. the SOFI $J$ filter.
The grisms 150I, 300I and 600B are described in the text. \\
  $^{(b)}$ Not corrected for Galactic foreground extinction.\\
 $^{(c)}$  Image quality affected by guiding problems. \\
  $^{(d)}$ Variable seeing due to varying cirrus. }
\end{table*}


\subsection{Spectroscopy}

Spectroscopic observations were performed during the first two days
in several wavelength bands to derive the redshift and during the
maximum of the late-time bump to search for signatures
of the potential supernova and to derive host galaxy parameters.
In the $JK$ bands the short-wavelength arm of ISAAC was used at
low resolution (with a final dispersion of 3.6 \AA/pixel in the $J$ band, 
and 7.2 \AA/pixel in the $K$ band), 
The 1\farcs0 slit was used.
The $JK$ band spectra were both taken at 1\farcs2 seeing, thus giving 
a FWHM resolution of 29 \AA\ ($J$ band) and  59 \AA\ ($K$ band).

Optical spectra were taken with FORS2 at the Yepun telescope at three
different occasions, each with a different grism (see Tab. \ref{log}).
The 150I grism 
has a mean
 5.5\AA/pixel scale, which at the 0\farcs5 seeing and the use of a 
0\farcs7 slit led to a resolution of 14 \AA\ (FWHM).
Grism 300I 
has a  mean
2.59 \AA/pixel scale, leading to a resolution of 13 \AA\ (FWHM) at
0\farcs7--1\farcs2 seeing and use of the 1\farcs0 slit.
Finally,  grism 600B 
with a  mean
1.2 \AA/pixel scale at 1\farcs2--1\farcs5 seeing and the 1\farcs0 slit
led to a FWHM resolution of 6 \AA.
The pixel scale changes by less than 5\% from the red to the blue end
of each grism. The peak spectral response for the 150I, 300I and 600B
grism is at 5000 \AA, 7500 \AA\ and 4000 \AA, respectively.


\begin{table}[ht]
\caption{Local photometric $B, V$ standards}
\hspace{-0.5cm}
\begin{tabular}{lccc}
   \hline \noalign{\smallskip} Star & Coordinates (J2000.0) & $B$ &
   $V$ \\ \noalign{\smallskip} \hline \noalign{\smallskip} 
   B1 & 11\h34\m18\fss9 --76\degr01\amin37\asec & 20.11$\pm$0.01 &
   18.93$\pm$0.01   \\ 
   B2 & 11\h34\m19\fss7
   --76\degr02\amin06\asec & 20.95$\pm$0.01 & 19.95$\pm$0.02   \\ 
   B3 & 11\h34\m27\fss2 --76\degr01\amin52\asec& 22.93$\pm$0.03 &
   21.93$\pm$0.04   \\ 
   B4 & 11\h34\m27\fss2
   --76\degr02\amin34\asec & 23.94$\pm$0.07 & 22.49$\pm$0.05   \\ 
   B5 & 11\h34\m27\fss4 --76\degr02\amin22\asec & 21.58$\pm$0.01 &
   20.35$\pm$0.02   \\ 
   B6 & 11\h34\m29\fss9
   --76\degr02\amin36\asec& 23.69$\pm$0.05 & 22.27$\pm$0.05   \\
   B7 & 11\h34\m33\fss3 --76\degr01\amin33\asec& 20.64$\pm$0.01
   & 19.59$\pm$0.01   \\ 
   B8 & 11\h34\m36\fss2
   --76\degr01\amin07\asec& 23.00$\pm$0.05 & 21.41$\pm$0.04   \\ 
   B9 & 11\h34\m38\fss9 --76\degr01\amin38\asec& 20.37$\pm$0.01 &
   19.03$\pm$0.01   \\ 
   B10 & 11\h34\m41\fss9
   --76\degr01\amin04\asec& 25.25$\pm$0.16 & 23.01$\pm$0.09   \\
   B11 & 11\h34\m52\fss9 --76\degr01\amin31\asec& 23.50$\pm$0.05 &
   21.57$\pm$0.03   \\ 
   \noalign{\smallskip} \hline
   \label{bstand}
\end{tabular}

\noindent B4=R5, B6=R6, B7=J2.
\end{table}

\begin{table}[ht]
\caption{Local photometric $R_{\rm c}, I_{\rm c}$ standards}
\hspace{-0.5cm}
\begin{tabular}{lccc}
   \hline \noalign{\smallskip} Star  & Coordinates (J2000.0) & $R_{\rm c}$ &
   $I_{\rm c}$ \\ \noalign{\smallskip} \hline \noalign{\smallskip} 
  R1 & 11\h34\m10\fss8 --76\degr02\amin50\asec& 20.87$\pm$0.05 &
   20.06$\pm$0.02   \\ 
  R2 & 11\h34\m11\fss2
   --76\degr03\amin23\asec& 20.96$\pm$0.02 & 20.35$\pm$0.02   \\ 
  R3 & 11\h34\m15\fss4 --76\degr00\amin37\asec& 19.97$\pm$0.04 &
   19.34$\pm$0.01   \\
  R4 & 11\h34\m18\fss0
   --76\degr02\amin32\asec& 23.43$\pm$0.10 & 22.95$\pm$0.08   \\
  R5 & 11\h34\m27\fss2 --76\degr02\amin34\asec& 21.75$\pm$0.04
   & 21.09$\pm$0.03   \\
  R6 & 11\h34\m29\fss9
   --76\degr02\amin36\asec& 21.48$\pm$0.04 & 20.80$\pm$0.03   \\
  R7  & 11\h34\m33\fss3 --76\degr02\amin30\asec& 21.82$\pm$0.06 &
   21.13$\pm$0.03   \\
  R8 & 11\h34\m40\fss8
   --76\degr02\amin36\asec& 20.47$\pm$0.03 & 19.98$\pm$0.01   \\
  R9  & 11\h34\m41\fss9 --76\degr01\amin35\asec& 22.36$\pm$0.05 &
   21.84$\pm$0.05   \\
  R10& 11\h34\m43\fss8
   --76\degr01\amin27\asec& 22.78$\pm$0.07 & 21.49$\pm$0.04   \\
  R11& 11\h34\m45\fss4 --76\degr01\amin32\asec& 19.36$\pm$0.08 &
   18.87$\pm$0.01   \\ 
\noalign{\smallskip} \hline
 \label{rstand}
\end{tabular}

\noindent R5=B4, R6=B6.
\end{table}

\begin{table*}
\caption{Local photometric standards for the NIR bands (Fig.~\ref{jfc})}
\begin{tabular}{lcccc}
   \hline \noalign{\smallskip} Star & Coordinates (J2000.0) & $J_{\rm s}$ &
   $H$ & $K_{\rm s}$ \\ \noalign{\smallskip} \hline \noalign{\smallskip}
  J1 &   11\h34\m39\fss0 --76\degr01\amin17\asec &  15.21$\pm$0.05 &
   14.81$\pm$0.05 & 14.91$\pm$0.05 \\
  J2 & 11\h34\m33\fss3
   --76\degr01\amin33\asec &  17.28$\pm$0.05 & 16.74$\pm$0.05 &
   16.82$\pm$0.05 \\
  J3  & 11\h34\m25\fss8 --76\degr01\amin37\asec
   &  18.56$\pm$0.05 & 18.28$\pm$0.05 & 18.27$\pm$0.05 \\
  J4  &
   11\h34\m35\fss6 --76\degr01\amin53\asec &  17.15$\pm$0.05 &
   16.31$\pm$0.05 & 16.31$\pm$0.05 \\
 \noalign{\smallskip} \hline
\end{tabular}
\label{jstand}
\end{table*}


\begin{figure*}
\begin{tabular}{c}
\psfig{figure=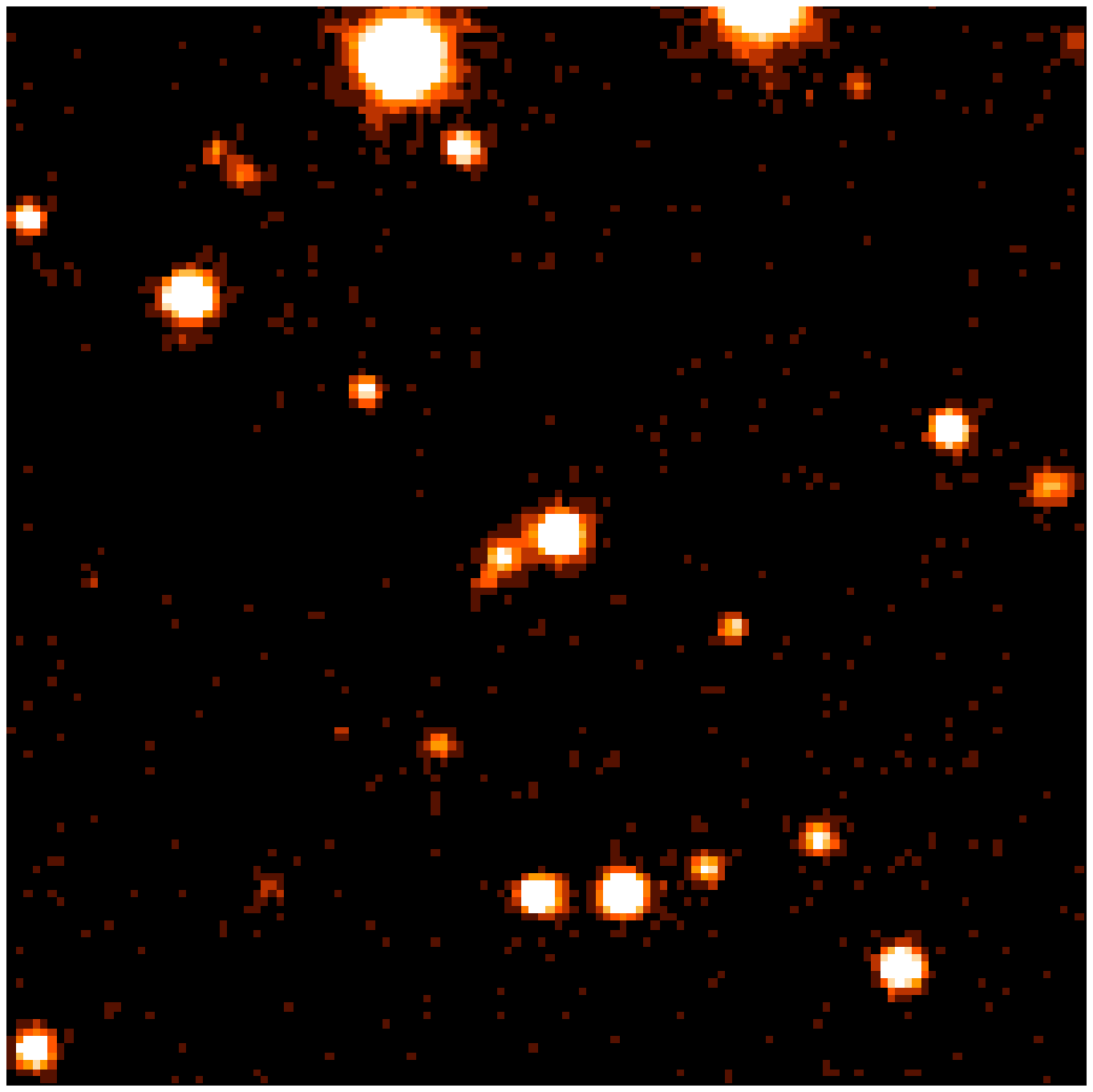,width=6.0cm,%
     bbllx=7.cm,bblly=15cm,bburx=13.5cm,bbury=21.5cm,clip=}
\psfig{figure=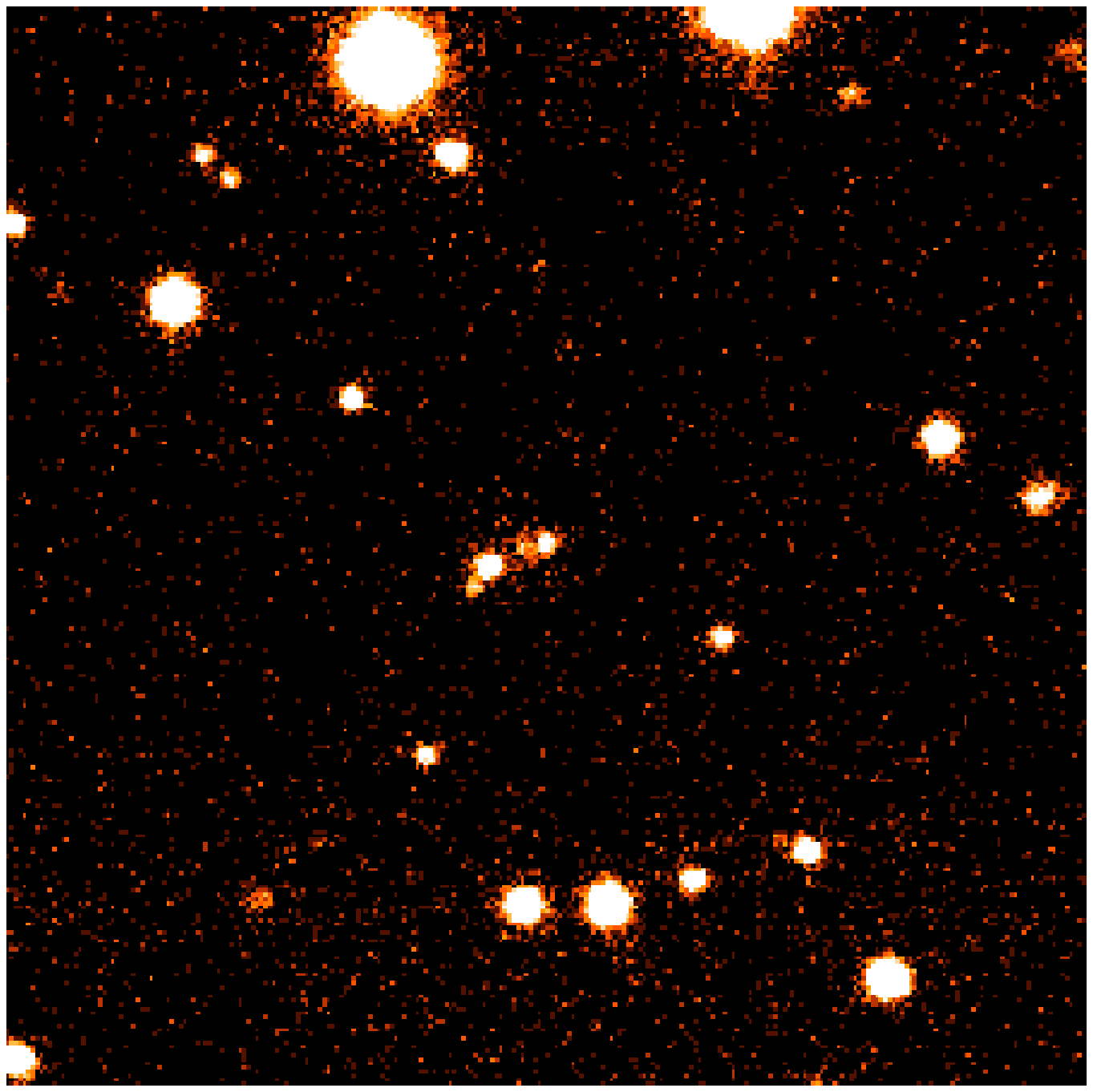,width=6.0cm,%
     bbllx=7.cm,bblly=15cm,bburx=13.5cm,bbury=21.5cm,clip=}
\psfig{figure=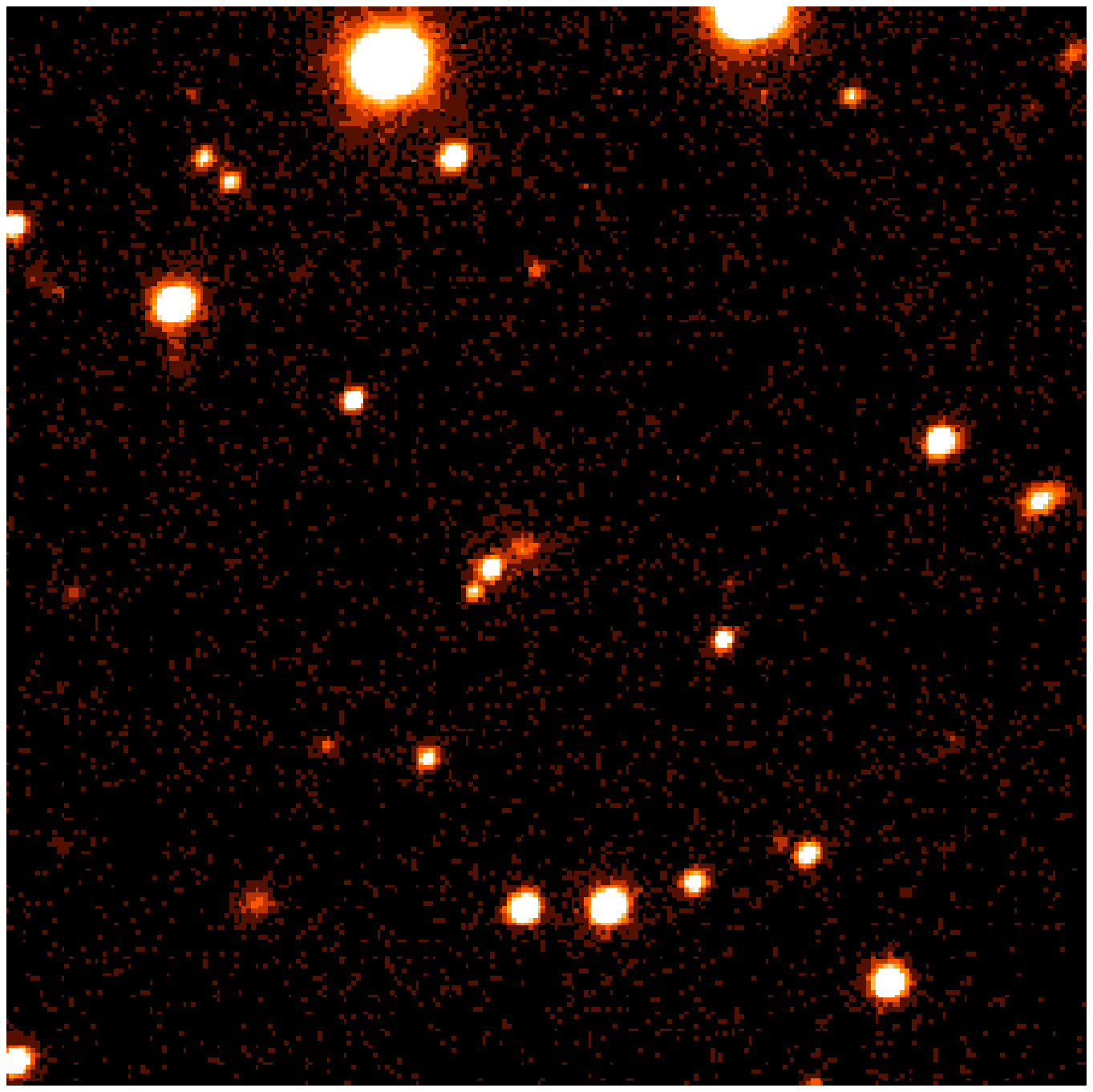,width=6.0cm,%
     bbllx=7.cm,bblly=15cm,bburx=13.5cm,bbury=21.5cm,clip=}\\
\end{tabular}
\caption[3im]{Sequence of $J$ band images of GRB 011121 taken with
   SOFI/NTT on 22 November 2001 (left), and with VLT/ISAAC on
   24 November 2001 (middle) and 9 February 2002 (right). 
   The right-most image clearly shows the host galaxy (labeled \#3 in 
   Fig. \ref{jfc}), whereas the GRB afterglow has disappeared. 
   The field is 17\asec$\times$17\asec, North at the top and East to the left.}
\label{3ima}
\end{figure*}


\section{Data reduction, analysis, and basic results}

\subsection{Photometry}

The optical and near-infrared images were reduced in standard fashion 
using IRAF\footnote{IRAF is distributed by the National Optical Astronomical
Observatories, which  is operated  by the Associated  Universities for
Research in  Astronomy, Inc., under  contract to the  National Science
Foundation.}
 as well as ESO's {\it eclipse} package (Devillard
2002). Photometric calibration of the GRB field
was performed using SExtractor (Bertin \& Arnouts 1996).
Airmass correction was done according to the coefficients
provided by ESO's Web pages (in units of mag airmass$^{-1}$):
$k_B=0.240\pm0.007$,
$k_V=0.112\pm0.005$,
$k_R=0.091\pm0.007$, 
$k_I=0.061\pm0.006$ for VLT FORS1 and FORS2, and
$k_J$=0.06,
$k_H$=0.06,
$k_K$=0.07 for VLT ISAAC at Paranal,
$k_B$=0.20,
$k_V$=0.11,
$k_R$=0.05,
$k_I$=0.02 for La Silla.
For the optical observations, the Landolt standard fields observed were
SA~92-249, MarkA, Rubin~152, GD 108 and SA~100. 
For the NIR calibration the UKIRT infrared standard star FS~12 was used.

Local photometric standards (Tables \ref{bstand}, \ref{rstand}, \ref{jstand}) 
were selected according to
their detection in $BVRI$ by SExtractor and a small scatter in their measured
magnitudes on the images taken at different observing epochs.
Moreover, care was taken that the difference between the deduced
Bessel-$R$ band magnitudes (FORS1) and the $R_{\rm special}$ magnitudes
(FORS2) of these stars was less than about 0.1 mag.
These local photometric standards were then used to derive the
magnitude of the optical transient (Tab. \ref{log})
after removal of the host galaxy.


\begin{figure*}[ht]
   \begin{tabular}{cc}
     \psfig{figure=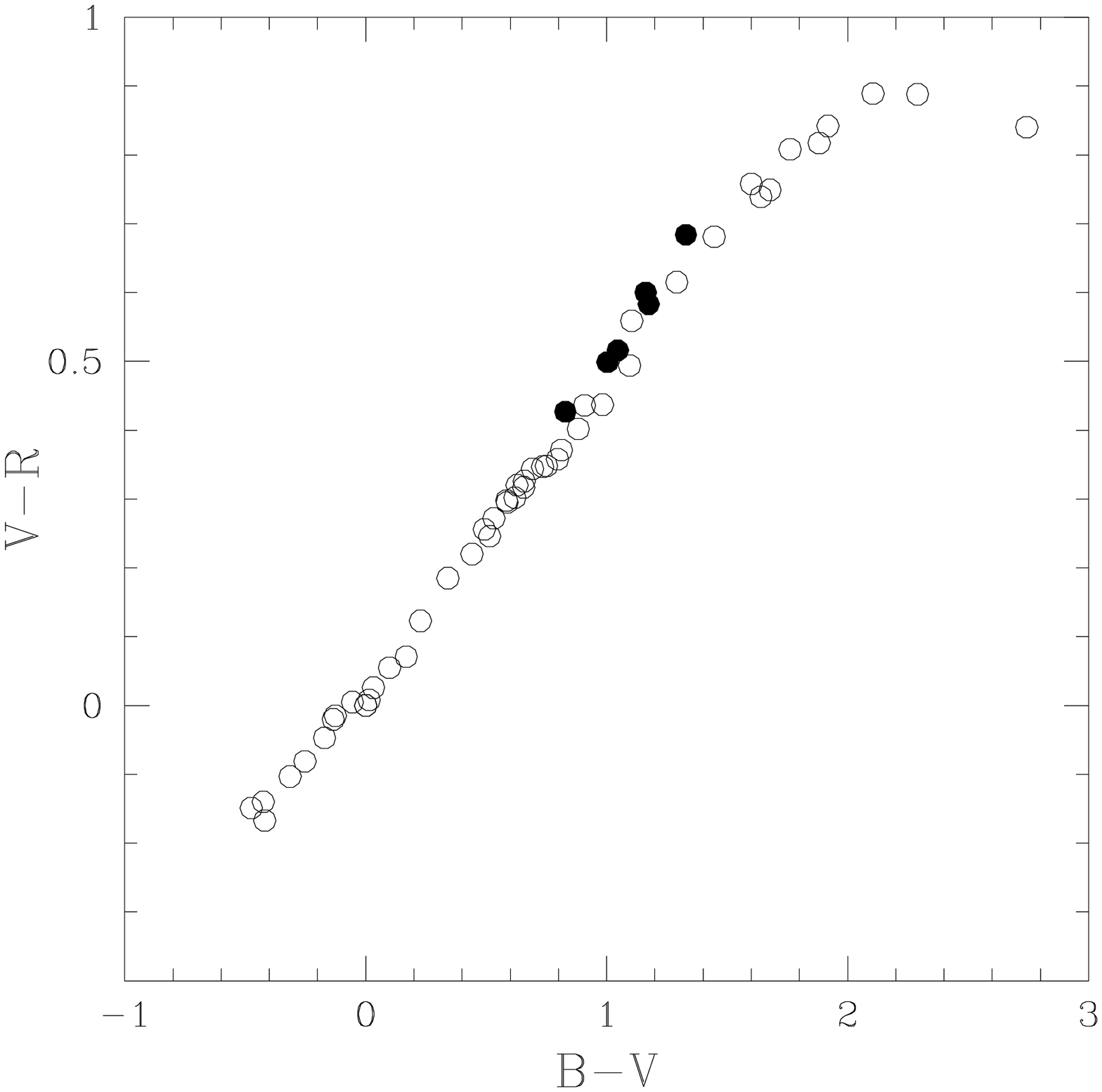,width=6.cm} &
     \psfig{figure=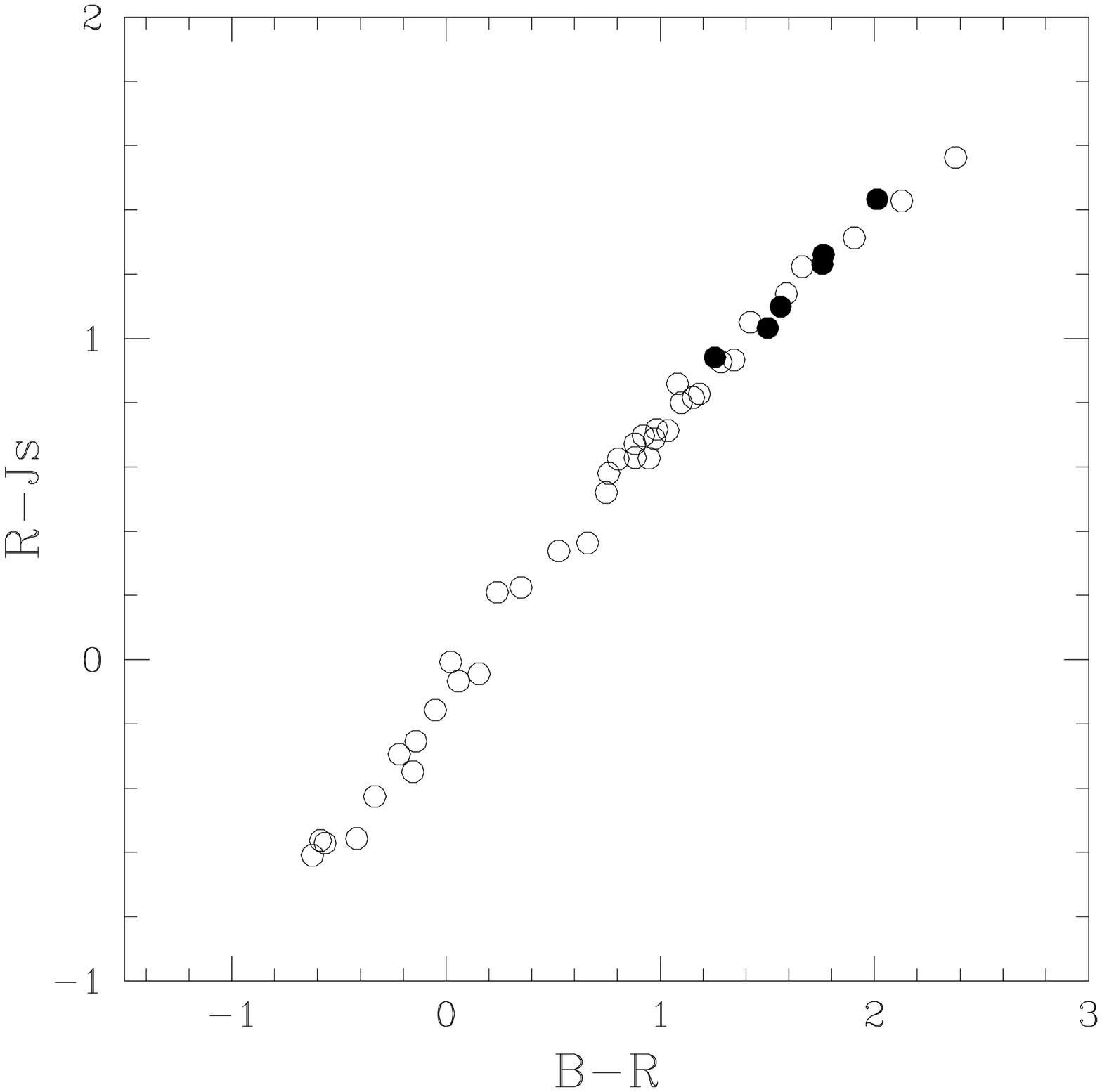,width=6.cm} \\
     \psfig{figure=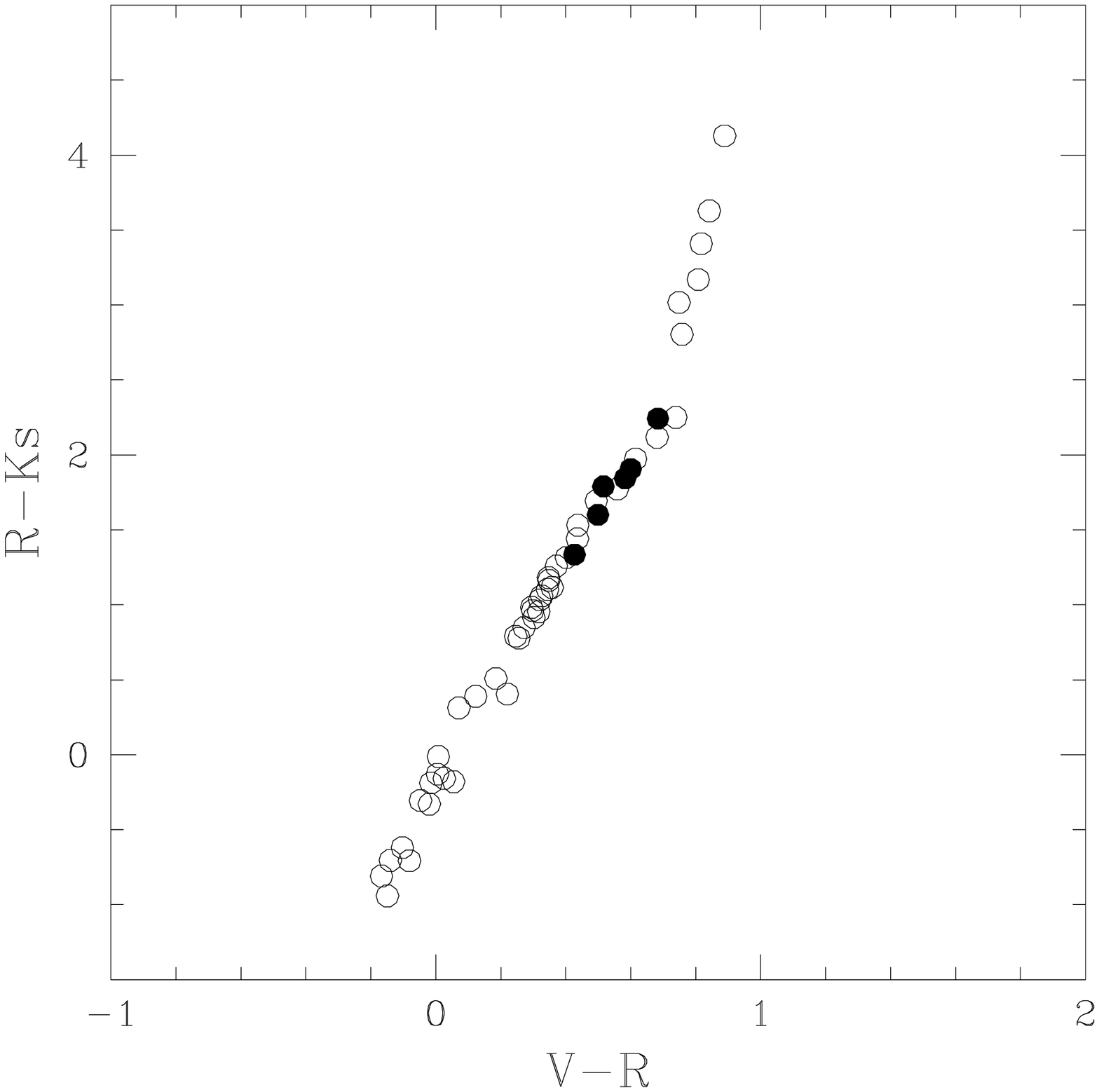,width=6.cm} &
     \psfig{figure=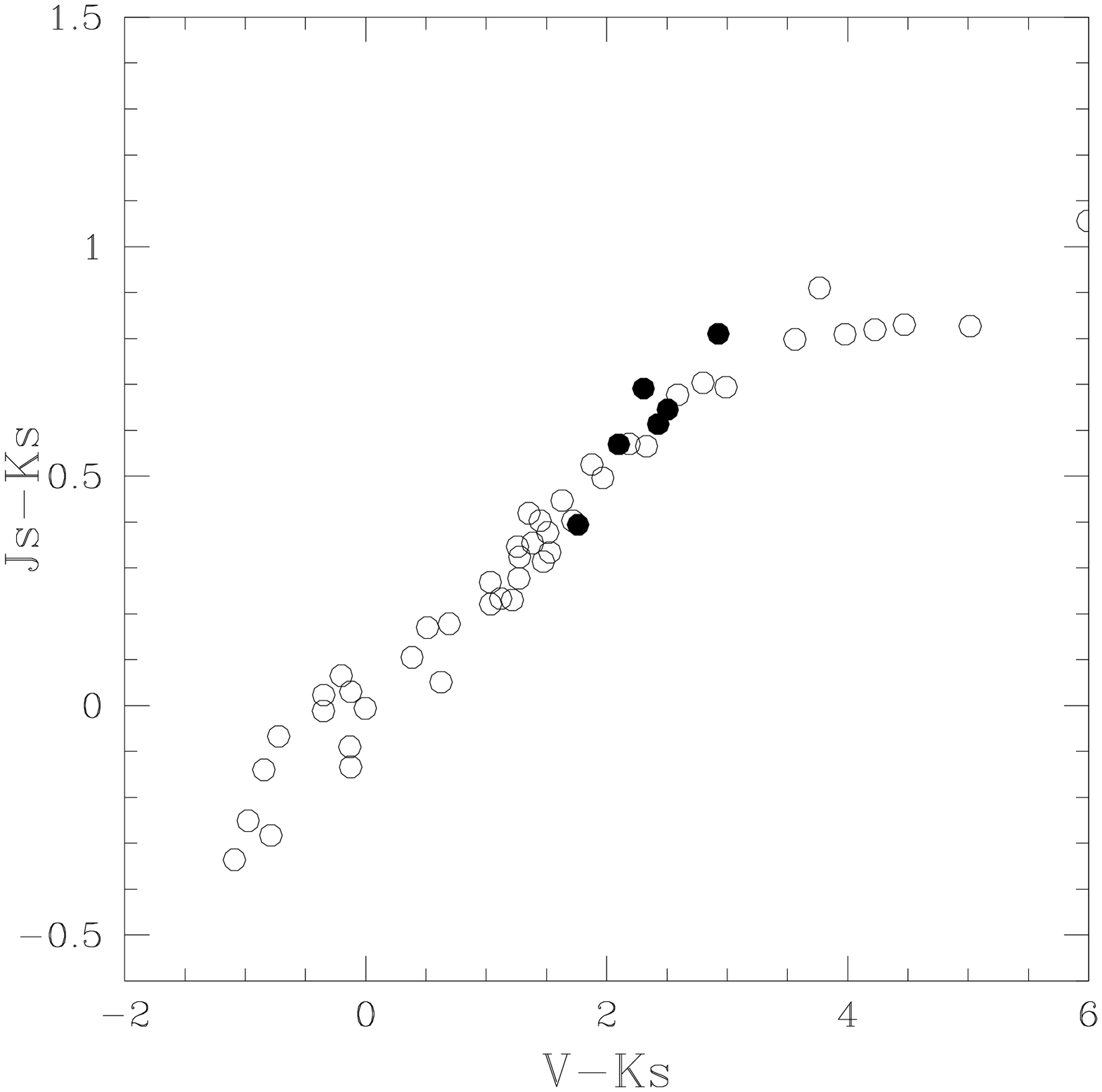,width=6.cm} \\
   \end{tabular}
\caption[stan]{Color-color diagrams of bright, non-saturated stars 
  detected in the central area  of the images of GRB 011121 (filled circles), 
  compared with the expected color-color sequence as derived from the 
  convolution of the stellar library of  Pickles with the
  total filter efficiency curves of the instruments used (open circles).
\label{stancol}}
\end{figure*}


The accuracy of the photometry in the different filters has been 
cross-checked on the set of images taken on 24 November 2001
via a comparison of the observed stellar brightnesses
versus those of expected synthetic stars.
For this we have first measured the magnitudes of our local standard
stars using YODA (Drory 2003), after having convolved the images to the 
common worst seeing of 1\farcs4. In a second step, we took
synthetic stellar spectra from the Pickles spectral library
and convolved those with the filter transmission curves and efficiencies
of the corresponding instrument (FORS2+ISAAC). Finally, we overplotted
the measured standards over the synthetic stars in various 
color-color diagrams (Fig. \ref{stancol}).
This shows that the photometric calibration is of coherent quality over
all the 7 filter bands used.

\subsection{The light curve \label{COBE} }

GRB 011121 occurred at galactic coordinates  $(l_{\rm II}, b_{\rm II}) =
297\fdg77, -12\fdg43$. The Schlegel, Finkbeiner, \& Davis (1998) extinction
maps predict $E(B-V)=0.46$ mag along this line of sight through the
Galaxy. Assuming the usual ratio of visual-to-selective extinction of 3.1,
this gives $A_V\approx$ 1.4 mag. The HI maps of Dickey \& Lockman (1990) give
$N_{\rm H} = 1.2 \times 10^{21}$ cm$^{-2}$ which translates into $A_V\approx$
0.9 mag using the extinction to absorption correlation of Predehl \& Schmitt
(1995).  We used the higher of the above two values, and then followed
standard procedures (Rieke \& Lebofsky 1985; Cardelli, Clayton, \& Mathis
1989; Reichart 2001) to calculate the extinction in the other photometric
bands, resulting in: $A_B$ = 1.87 mag, $A_{R_c}$ = 1.15 mag, $A_{I_c}$ = 0.83
mag, $A_J$ = 0.40 mag, $A_H$ = 0.25 mag and $A_{K_s}$ = 0.17 mag.  In all the
light curve plots this extinction has been corrected for.

\begin{figure*}[ht]
 \begin{tabular}{c}
   \psfig{figure=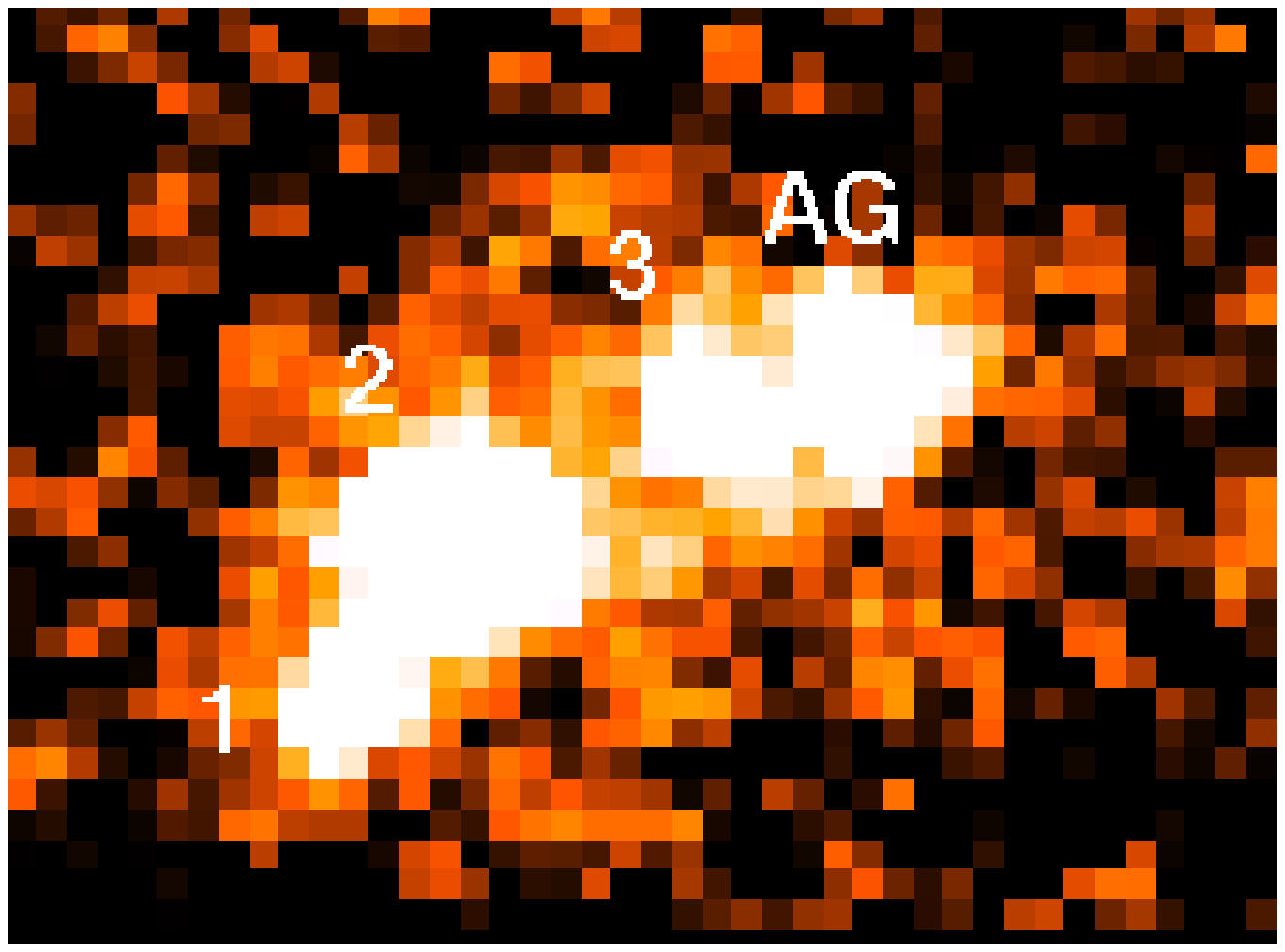,width=5.3cm} 
    \psfig{figure=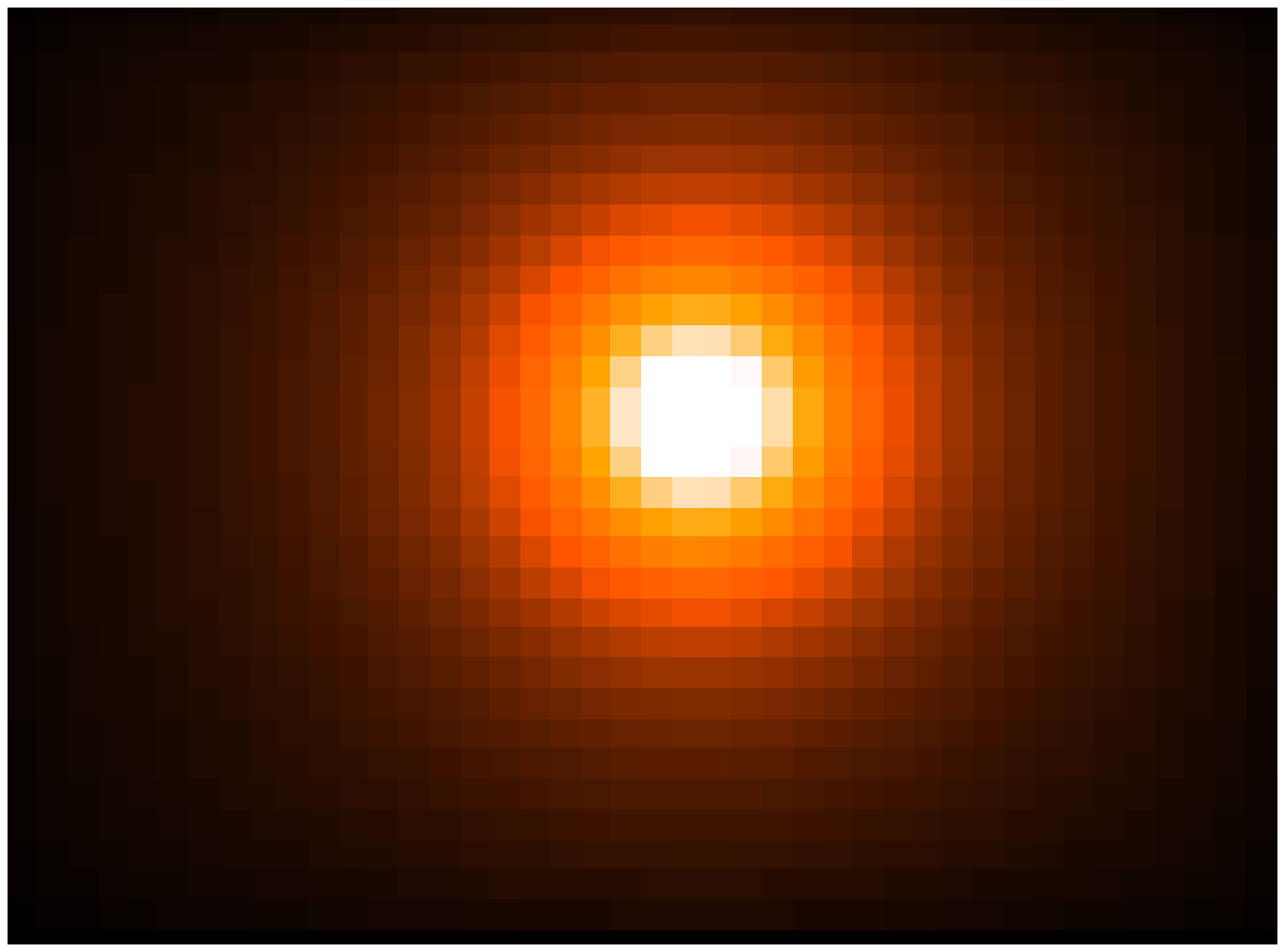,width=5.3cm}
    \psfig{figure=grb011121_subtrr.ps,width=5.3cm} \\
 \end{tabular}
  \caption[subtr]{Sequence of R band sub-images of 24 Nov. 2001
   showing the original (left), the model of the host galaxy (center)
   and the host-subtracted image (right). All sub-images are shown 
   at the same scale and with the same cut values. Labels are the same
   as in Fig. \ref{jfc}.  Photometry of the lower-left
   star in the original and the host-subtracted image has been
   done for all epochs to ensure that no oversubtraction occurred.
   The residual scatter in this comparison star was in the range of 
  0.02--0.05 mag.
\label{subtr}}
\end{figure*}

Accurate photometry of the afterglow requires a careful removal of
contaminating light from the underlying GRB host galaxy which has an angular
extent of two arcseconds (Fig. \ref{3ima}). This was accomplished with
a two-component intensity profile model, as described in more detail in
\S \ref{hostgalaxy}. The model is derived from the J band image
of 9 February 2002, taken at a seeing of 0\farcs45. We assume that the
radial profile of the host galaxy is the same in all filter bands,
as justified by a comparison of the scale lengths of 86 face-on galaxies
in the $B,V,R,H$ filter bands by de Jong (1996), who finds that
they are identical within the errors (the disk scale length ratio
of $R$ vs. $H$ band is 1.07$\pm$0.08).
The subtraction is done in count space on the images before
the photometry, and for each image the host model is first convolved
with the seeing of that exposure, and then iteratively adjusted in
intensity until the area around the afterglow is best matched to the
undisturbed background somewhat further away. As a consistency test
we computed the host magnitude for each image taking into account
the seeing convolution and the final intensity normalization, and
found that the host magnitude differs at most by 0.07 mag (two instances)
from the values given in \S \ref{hostgalaxy}. In addition, star 1 from Fig.
\ref{jfc} has been used to monitor the quality of the subtraction,
which was always accurate to better than 0.05 mag. This ensures that
the light curve shape is not contaminated by this subtraction procedure.
An example of the result is shown in Fig. \ref{subtr}.


\subsubsection{Light curve based on observations prior to t = 10 days 
               \label{ohneSN} }

\begin{figure}[ht]
\vbox{\psfig{figure=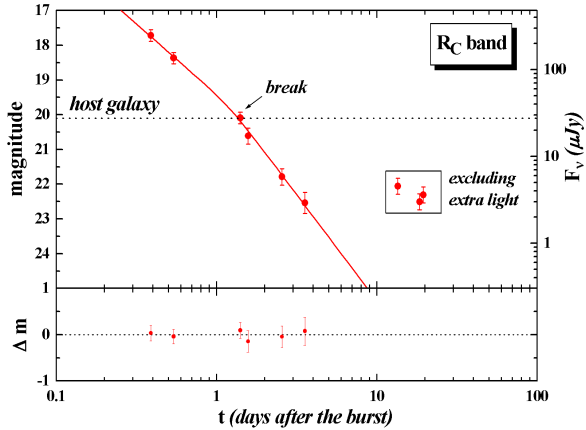,width=8.0cm,angle=0}}
\caption[bspec]{$R_c$-band light curve of the afterglow of GRB 011121
   and a fit to the data prior to $t$ = 10 days.  Compared to Tab. \ref{log}a
   conservative systematic  1$\sigma$ error of 0.15 mag has been added to
   all data. The break in the power-law 
   decay occurs at $t_b$ = 1.2 days. At late times the afterglow
   is significantly brighter than predicted by the power-law.
   Note the particularly bright host galaxy.}
\label{noSNfigure}
\end{figure}

Figure \ref{noSNfigure} shows the afterglow light curve in the $R_c$-band 
after removal of the flux from the underlying host galaxy, and after
correcting for galactic extinction. Theoretical fits were obtained with 
the formula of Beuermann et al. (1999) in the representation of  
Rhoads \& Fruchter (2001):
\begin{equation}
   F_\nu (t) = 2^{1/n} \ F_{\nu}(t_b)[(t/t_b)^{\alpha_1\,n}+
                    (t/t_b)^{\alpha_2\,n}]^{-1/n} \,.
\label{beuermann}
\end{equation}
Here $F_\nu$ is the flux density, $t_b$ the break time (in days), and $n$ the
parameter which describes the smoothness of the break. In all cases we fitted
apparent magnitudes. Here and in the following we use the standard notation
for the time and frequency dependence of the flux density in the simple
fireball model: $F_\nu \propto t^{-\alpha}\, \nu^{-\beta}$.

The fit of the $R_c$-band data excluding the measurements at $t\ge10$ days
gives an early-time slope of $\alpha_1=1.62\pm0.62$, a late-time slope of
$\alpha_2 = 2.44\pm 0.38$, and a break time $t_b = 1.2 \pm 1.0$ days
($\chi^2/$d.o.f. = 0.45; d.o.f. $\equiv$ degrees of freedom). 
The parameter $n$ which measures the sharpness 
of the break, was fixed at $n$ = 10, 
corresponding to a sharp break. 
However, values as low as $n$ = 1 give similarly good fits. 

At early times (22 November 2001, 3:30 UT)  we measure
$R_{\rm c}-K = 2.7\pm 0.1$ mag 
(after correction for galactic extinction). Hence, the spectral slope of the
afterglow is $\beta = 0.80\pm 0.15$.
Two days later, on 24 November 2001, 7:00 UT,
we find $\beta = 0.62\pm 0.05$ based on our $BVRJHK$ data,
which is consistent with the earlier result, and which is fully
in agreement with the results obtained by
Garnavich \etal\ (2003) and Price \etal\ (2002).
These afterglow parameters
are fully consistent with those expected from the simple versions of the
fireball as well as observations of previous gamma-ray burst afterglows (e.g.,
van Paradijs, Kouveliotou, \& Wijers 2000).

We can improve the fit by adding more data points as follows.  First, using
the deduced spectral index of the early afterglow we can estimate  that at the
time of the single AAO observations (Table~\ref{log}), i.e.  before the
deduced break time, the  $R_c$-band magnitude of the optical transient was
20.12$\pm$0.25. Second, we can include in our fit the $R$-band data from
Garnavich et al. (2003) for $t<0.6$ days. In doing so, we get
$\alpha_1=1.62\pm0.39$, $\alpha_2 = 2.44\pm 0.34$, $t_b = 1.20 \pm 0.75$ days,
$\chi^2/$d.o.f. = 1.11.

\subsubsection{Light curve based on all observations \label{mitSN} }

Including the ``late-time" data (after $t$ = 10 days) to the fits
adds substantially more freedom, since different model components
can now compensate each other. As before,
we only use $R_c$-band data for the analysis.  In order to account for the
excess light after $t$ = 10 days, we assume that this ``bump'' is due to
light from an underlying supernova. 
In fact, it has been officially designated 
as SN 2001ke (Garnavich \etal\ 2003).
To model this supernova component we
employ the observed $UBVR_cI_c$ light curves of SN 1998bw (Galama et al. 1998)
as a template. 

\begin{figure}[ht]
\vbox{\psfig{figure=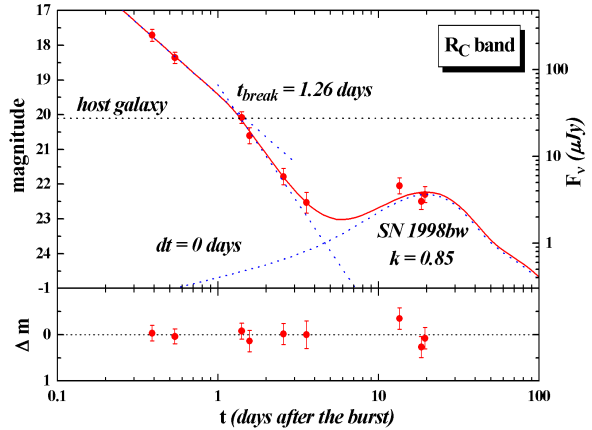,width=8.0cm,angle=0}}
\caption[bspec]{$R_c$-band light curve based on 
   eq.~(\ref{beuermann}) and an additional supernova component, assuming
   $dt$ = 0 days (eq.~\ref{dt}).  Note that all photometric data are
   corrected for Galactic extinction (\S \ref{COBE}).
   Compared to Tab. \ref{log} a
   conservative systematic  1$\sigma$ error of 0.15 mag has been added to
   all data.}
\label{Rcbandfit0}
\end{figure}

\begin{figure}[ht]
\vbox{\psfig{figure=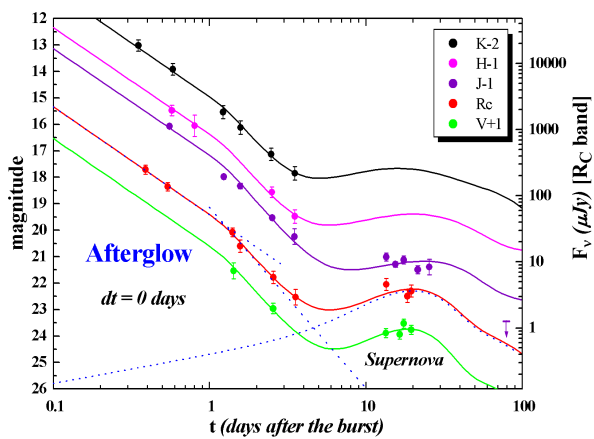,width=8.0cm,angle=0}}
\caption[bspec]{The same as Fig.~\ref{Rcbandfit0} but now showing
   the best fit in all photometric bands. When fitting the  $V, J, H,
   K_s$-band light curves the afterglow parameters deduced from the
   $R_c$-band fit ($\alpha_1, \alpha_2, [n], t_b$) were used as 
   input, i.e. the functional form of the afterglow
    light curve was fixed. The supernova light curve (extrapolated towards the
   $H$ and $K$ bands), however, was
    calculated for the chosen photometric band.
    }
\label{Allbandfit0}
\end{figure}

The effects of the redshift of GRB 011121 ($z$=0.36) were taken
into account assuming the cosmological parameters as given in section 
\ref{gamma}.
For SN 1998bw we used a redshift of $z = 0.0085$ (Tinney \etal\ 1998).
The entire numerical procedure is in detail explained in Zeh \& Klose
(2003). It interpolates smoothly between the $UBVRI$ light curves of
SN 1998bw (Galama \etal\ 1998)\footnote{see also 
http://zon.wins.uva.nl/~titus/grb980425.html},
so that for any given
time a set of flux densities is calculated. The predicted
time-dependent supernova light curve of a redshifted GRB-supernova
based on the SN 1998bw  template is then calculated according to the
procedure outlined by Dado \etal\ (2002a).  The apparent
magnitude of a redshifted SN 1998bw in a given photometric band is
finally obtained by integrating over the flux densities (transformed
into per unit wavelength), multiplied by the corresponding filter
response function, and by applying the usual normalization
factors. The entire procedure has been succesfully applied for GRB
030329/SN 2003dh (Hjorth \etal\ 2003; Zeh \etal\ 2003) and has been
successfully tested against the results obtained by others for other
bursts (e.g., Masetti \etal\ 2003; Dado \etal\ 2002a).
 The results are shown in
Figs.~\ref{Rcbandfit0} and \ref{Allbandfit0}.


The resulting parameters are: $\alpha_1 = 1.63 \pm 0.61, \alpha_2 = 2.73 \pm
0.45,  t_b = 1.26 \pm 0.94$ days and $k = 0.85 \pm 0.11$ with
$\chi^2/$d.o.f. = 1.11. Here, $k$ is the luminosity ratio in the $R_c$ band
between the GRB 011121 supernova and SN 1998bw at maximum (at $z=0.36$). As
before, the parameter $n$ was held constant at  10 because  otherwise
the sparse data would not allow convergence of the numerical fitting
algorithm. Our conclusions are not sensitive to the exact value of this
parameter.
On the other hand, we emphasize that the value we get for the
parameter $k$  assumes a 'perfect' SN 1998bw light curve at the redshift of the
burster. Any potential relation between luminosity and light curve
shape of a supernova in a certain photometric band is neglected here.
In particular, $k$ does not say anything about the bolometric
luminosity of the GRB supernova compared to SN 1998bw.
This parameter is also sensitive to potential additional parameters
which one can introduce in order to improve the fit of the SN light
curve (see \S~\ref{grbsncon}).

For SN 1998bw we assumed zero extinction along the line of sight 
through our Galaxy. Accordingly, our results obtained for the 
luminosity of the supernova accompanying GRB 011121 
in units of the luminosity of SN 1998bw, the parameter $k$, scale as  
$k\rightarrow k\,\times\,exp(-A_V(\mbox{SN 1998bw})/1.086)$, if such an 
extinction is taken into account. Note also that we assumed  a 
Galactic extinction along the line of sight of $A_V$(Gal)= 1.4 mag. 
If smaller values are preferred (Bloom \etal\ 2002), 
$k$ has to be corrected/reduced again.

Figures \ref{3ima} (middle panel), \ref{Rcbandfit0} and \ref{Allbandfit0} 
show that the SN
light already affected the afterglow light a few days  after the GRB. This
implies a substantial impact on the measurement  of the parameter
$\alpha_2$. Unfortunately, due to instrumental constraints, we have a data gap
between days 4 and 10 after the burst trigger so that the 1$\sigma$ error of
the deduced $\alpha_2$ rises substantially when we include the SN light in the
fit. Note that most of the data published previously are not host subtracted,
and therefore cannot be simply added to our data set (Brown \etal\ 2001;
Phillips \etal\ 2001; Price \etal\ 2001; Stanek \& Wyrzykowski 2001).  Despite
this added uncertainty, our analysis clearly indicates the presence of a break
about one day after the burst. The exact  time of the break is sensitive to
the details of the fit (compare the results we obtain by fitting data prior to
$t$ = 10 days vs. fitting all the data), but the existence of the break at
$t\sim$ 1 day is clearly established. Previous studies of the afterglow light
curve  of GRB 011121 did not find this break (e.g., Price et al. 2002;
Garnavich et al. 2003), which  can be attributed to sparse sampling of the
afterglow at this particular time. It remains difficult to explain why
the radio data of the afterglow do not agree with a jet model 
(Price \etal\ 2002). 


\subsection{The spectrum of the afterglow}

\begin{figure}[ht]
\vbox{\psfig{figure=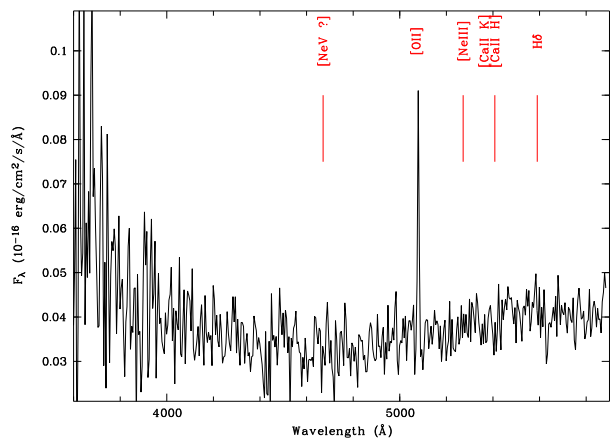,width=8.cm,angle=270,%
          bbllx=1.6cm,bblly=2.5cm,bburx=20.5cm,bbury=27.2cm,clip=}}\par
\caption[bspec]{Spectrum of GRB 011121 in the blue band, taken on 23
    Nov. 2001 with VLT/FORS2 equipped with the 600B grism. 
  While only the [O~II] line is detected beyond doubt, we also indicate
  the positions of other lines which could have been expected.}
\label{Bspec}
\end{figure}

The VLT/FORS2 spectrum of the afterglow (including light from the host galaxy
and the underlying supernova) contains
several strong emission lines  (Figs.~\ref{Bspec}, 
Tab.~\ref{lflux}), but no absorption lines. The redshift determined from 
these lines is $z$ = 0.362$\pm$0.001 (see the figures for line
identifications), consistent with results of Infante
\etal\ (2001). This redshift corresponds to a luminosity distance of 2.07 Gpc
(assuming the cosmological parameters of section~\ref{gamma}), and a distance 
modulus of 41.59 mag.

Using the foreground $A_{\rm V}$ (\S 3.2), we have determined the extinction 
values in the lines $A^{\rm line}$, and corrected the measured line fluxes 
according to $F_{\rm corr}$ = $F_{\rm obs}$ $\times$ exp($\tau$), where
 $\tau$ = 1/1.086 $\times$ $A^{\rm line}$ is the continuum extinction
according to the galactic  $A_{\rm V}$.


\begin{table*}
\caption{Measured line fluxes (corrected for galactic extinction) and 
luminosities. \label{lflux}}
\begin{tabular}{ccccc}
   \hline \noalign{\smallskip} 
  Line & $\lambda_{\rm obs}$ & $z$ & Flux & Luminosity   \\ 
       &  (\AA)           &  & (10$^{-16}$ erg/cm$^2$/s) & (10$^{40}$ erg/s)\\ 
 \noalign{\smallskip} \hline
   \noalign{\smallskip} 
  \[OII] 3726  &  5079 & 0.363 & 0.41$\pm$0.03 & 8.6 \\ 
  H$\beta$            &  6625 & 0.362 & 0.22$\pm$0.04 / 0.14$\pm$0.04$^{(1)}$ & 3.2/2.0\\ 
  \[OIII] 4963 &  6760 & 0.362 & 0.11$\pm$0.06 / 0.08$\pm$0.05$^{(1)}$ & 1.6/1.1\\ 
  \[OIII] 5007 &  6821 & 0.362 & 0.35$\pm$0.05 / 0.16$\pm$0.04$^{(1)}$ & 4.9/2.2\\ 
  H$\alpha$           &  8945 & 0.363 & 0.96$\pm$0.08 / 0.81$\pm$0.09$^{(1)}$ & 9.2/7.7 \\ 
  Pa$\alpha$          & 25550 & 0.362 &  2.58$\pm$0.12 & 14.8\\ 
\noalign{\smallskip} \hline
\end{tabular}

\noindent{$^{(1)}$ The first number is measured from the 150I spectrum
on 25 November 2001, while the second number is measured from the 300I
spectrum on 12 December 2001.}
\end{table*}

\section{Results and Discussion}

\subsection{The afterglow model and the $\alpha-\beta$ relations}

Since our finding of a jetted explosion is crucial for an understanding of the
observational data we have carefully checked  this result. When we fit our
$R_c$-band data (Table~\ref{log}) with a single  power-law plus SN component
we get $\alpha = 1.98 \pm 0.11, k = 0.81 \pm 0.11$  with $\chi^2/$d.o.f. =
1.49. Not only is the fit worse in comparison to the fit based on the
Beuermann equation, the value we now  get for $\alpha$ is notably larger 
(3$\sigma$ deviation) than
the one deduced  for the early-time slope of the afterglow light curve (Price
et al. 2002; Garnavich et al. 2003). 
This again points to a change in the fading behavior
of the afterglow between about 0.4 and 2 days after the burst.  Moreover, our
deduced $\alpha_1$ for the early-time slope of the afterglow light curve is
fully consistent with  the result obtained by Price et al. (2002) and
Garnavich et al. (2003) based on their optical/NIR data. In fact, when we fit
their data we confirm that they do not show evidence for a break in the light
curve. We attribute this to the lack of data around $t=1$ day. This
makes us confident that the break in the light curve which we deduce 
based on our observations is a real effect.

In Tab. \ref{alphabeta} we list the predicted spectral slope $\beta$
corresponding to the measured $\alpha$'s for various afterglow
scenarios. Basically, we have to decide here between an ISM and a wind model
for a jetted explosion and to constrain the position of the cooling break
frequency in the spectral energy distribution (SED) of the afterglow light  
at the time of the observation. 
From the comparisons
of the predicted $\beta$, based on the theoretical $\alpha-\beta$ relations,
with the observed one on 22 November 2001, 3:30 UT, and on
24 November 2001, 7:00 UT, we conclude (see also Fig.~\ref{broadband}):

\begin{enumerate}
\item The model predictions agree best with the observations for the wind
      scenario. In this model, the presence of a late ``bump'' is
      the natural consequence of the supernova following the stage of rapid
      mass loss in a massive progenitor star.
\item 
The data favor the interpretation that during our observations the
      cooling break frequency $\nu_c$,
      which separates the contribution of fast-cooling electrons
      from slow-cooling electrons in the afterglow light
      (see Sari \etal\ 1998 for their original definition),
      was above the optical/NIR bands ($\nu_c
      \gr 10^{15}$ Hz). This is in agreement with the conclusions drawn by
      Price \etal\ (2002) based on their observational data; we
      refer the reader to this publication for a further discussion
      of this point.
\item The power-law index $p$ of the electron energy distribution,
      $N(\gamma_e)\, d\gamma_e \propto \gamma_e^{-p}\, d\gamma_e$
      was close to the expectations from particle acceleration
      in relativistic shocks (e.g., Sari \etal\ 1999).
\end{enumerate}

Within our measurement errors, there is no need for an
additional extinction by dust in the GRB host galaxy. A wavelength-dependent
extinction by cosmic dust in the GRB host galaxy would tend to increase
$\beta$, i.e., redden the afterglow light. For example, this was 
observed in the optical/near-infrared  afterglow of GRB 000418 (Klose
\etal\ 2000). The low value of $\beta$ we and others (Garnavich \etal\ 2003; 
Price \etal\  2002) find for the afterglow of GRB 011121
however, after correction for the influence of reddening by Galactic
dust, compared with various model predictions based on the observed
light curve shape (Tab.~\ref{alphabeta}), gives us no strong hint for
an additional dust component acting along the line of sight. Grey dust
in the GRB host galaxy could still be there (wavelength-independent
scattering cross section in the considered photometric bands), but this
cannot be deduced from our data.

On the other hand, one should be aware of the fact 
that the results obtained are sensitive to the adopted Galactic
extinction, $A_V$(Gal), along the line of sight. For example, if we
had used $A_V$(Gal) = 0.9 mag, as it suggested by the H~I maps
(\S~\ref{COBE}), then the extinction corrected $\beta$ for the
intrinsic optical afterglow would have been 1.1, based on our measured
$R_c-K$ color of the optical transient on 22 November 2001, 3:30
UT. We favor the higher extinction value (1.4 mag), however. The early
multi-color $UBVRIJK$ observations of the GRB afterglow by Garnavich
\etal\ (2002) give a best fit for the Galactic reddening along the
line of sight of $E(B-V)=0.43\pm 0.07$ mag. For a standard ratio of
total-to-selective extinction of 3.1, this is basically consistent with
our chosen value for $A_V$(Gal). We note also that an adopted
extinction of $A_V$(Gal) = 1.4 mag is still consistent with the result
deduced by Price \etal\ (2001) within their claimed 1$\sigma$ error bar
($A_V$(Gal) = 1.16 $\pm$ 0.25 mag).


\begin{table*}[t]
\caption{Predicted $\beta$ values for various afterglow scenarios.
Assuming a
relativistic jetted explosion then for observations at $t<t_{\rm break}$
(pre-break time) the isotropic model holds and $\alpha = \alpha_1$, whereas
for $t>t_{\rm break}$  (post-break time) the jet model applies and $\alpha =
\alpha_2$. We use here $\alpha_1 = 1.62 \pm 0.39$ and $\alpha_2 =
2.44 \pm 0.34$ (see \S~\ref{ohneSN}).
The parameter $s$ is the power-law index of the density profile of
the circumburst medium, $n(r) \propto r^{-s}$.  For an ISM model $s=0$, for a
wind model $s=2$. Case 1 stands for  $\nu > \nu_c$, case 2 for $\nu < \nu_c.$
In the former case the electron power-law index is given by $p=2\beta$,
whereas in the latter case $p=2\beta+1$ (e.g., Sari et al. 1999).}
\begin{tabular}{cllcc}
&&&&  \\[-2mm] \hline  &&&&  \\[-2mm]   
afterglow model & $\beta(\alpha)$ & predicted $\beta$ & electron $p$   \\[2mm]
 \hline
&&&& \\[-2mm]   
ISM,  iso, case 1 & $(2\alpha+1)/3$ & 1.41$\pm$0.26 & $2.82\pm$0.52\\ 
ISM,  iso, case 2 & $ 2\alpha/3   $ & 1.08$\pm$0.26 & $3.16\pm$0.52\\[1mm] 
ISM,  jet, case 1 & $\alpha/2$      & 1.22$\pm$0.17 & $2.44\pm$0.34\\ 
ISM,  jet, case 2 & $(\alpha-1)/2$  & 0.72$\pm$0.17 & $2.44\pm$0.34\\[1mm]
  \hline \\[-3mm] 
wind, iso, case 1 & $(2\alpha+1)/3$ & 1.41$\pm$0.26 & $2.82\pm$0.52\\ 
wind, iso, case 2 & (2$\alpha-1$)/3 & 0.75$\pm$0.26 & $2.50\pm$0.52\\[1mm]
wind, jet, case 1 & $\alpha/2$      & 1.22$\pm$0.17 & $2.44\pm$0.34\\ 
wind, jet, case 2 & $(\alpha-1)/2$  & 0.72$\pm$0.17 & $2.44\pm$0.34\\[1mm]
\hline
\end{tabular}
\label{alphabeta}
\end{table*}


We note that the observed change in the decay slope of the light curve around
the break time is in agreement with the predictions of the jet-wind model,
provided that the steepening of the light curve is due to the sideways
expansion of the jet (Rhoads 1999). 

It has been argued (Kumar \& Panaitescu 2000) that for an external density
profile as $r^{-2}$ the jet break in the light curve is expected to be very
gradual, taking at least two decades in time before most of the steepening
sets in. In the case of GRB 011121, the break in the light curve is
certainly shallower, extending no more than one decade in time
(even with fits using $n$=1 the ``break'' extends only from 0.5--2.5 days,
still less than a decade).
While this may be considered a problem, more extensive considerations
have shown a large diversity of light curve shapes, depending not only
on the density profile but also on the evolution of the Lorentz factor,
whether jets are uniform or non-uniform, and the difference of viewing
and jet angle (e.g., Wei \& Jin 2003).

\subsection{The supernova and the wind signature}

\subsubsection{Supernova features}

The light curve of GRB 011121 provides the clearest case to date for
excess emission above the usual power law extrapolation. The excess
in this case appears to become significant about one week after the burst
(Fig. \ref{Allbandfit0}). This feature has been
heralded as strong evidence for a supernova component (Bloom \etal\
2002; Dado \etal\ 2002b; Garnavich \etal\ 2003; Price \etal\ 2002).
Garnavich et al. (2003) proposed the supernova label SN 2001ke for
this afterglow, although no obvious supernova features were apparent
in their Magellan/LDSS2 spectrum obtained on Dec 7, 2001. While it
is far from proven that the late emission observed in GRB 011121 is
in fact due to a supernova, this interpretation is most natural. If 
we believe the SN-bump picture, the observations of Garnavich \etal\ (2003)
provide a stern warning: SN1998bw (used in our analysis) may not be 
appropriate as a ``template'' for SN-light associated with gamma-ray
bursts. Given that to date we have a rather limited sample of supernovae
associated with GRBs (see the recent review by Weiler \etal\ 2002), 
we are limited in what we can conclude from the bump in the afterglow
of GRB 011121, except that it is obvious that the simple afterglow
model is unable to explain features like these. Shock rejuvenation
and inhomogeneities in the GRB environment or energy injection can 
explain the kind of bumpy afterglow observed in the case of 
GRB 021004 (e.g., Lazzati \etal\ 2002, Holland \etal\ 2003), but does
not account for the long-term (weeks), sustained bump observed in
GRB 011121. Supernovae, on the other hand, provide a natural explanation
for the energy and also the time scale involved in the late bump
we are considering here.

\begin{figure}[ht]
\vbox{\psfig{figure=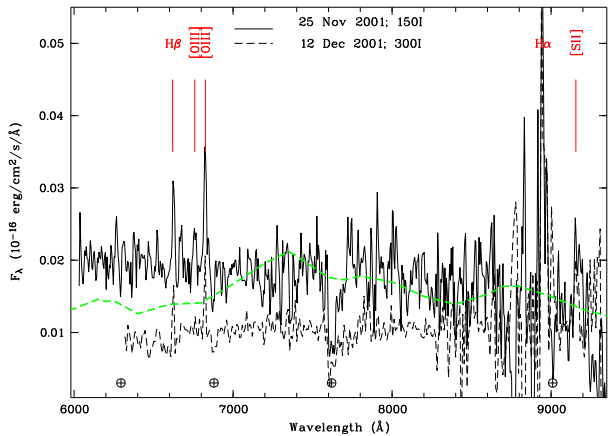,width=8.cm,angle=270,%
          bbllx=1.6cm,bblly=2.5cm,bburx=20.5cm,bbury=27.2cm,clip=}}\par
\caption[ispec]{Spectrum of GRB 011121 in the red band, taken with
 VLT/FORS2 equipped with the 150I (25 Nov. 2001, top) and  300I (12
 December 2001, bottom;  shifted by 0.01 units downwards) grisms,
 respectively.  Some line identifications are indicated.  Taking slit
 losses on 12 December 2001 into consideration, both
 spectra represent basically the host spectrum,
 since at the covered wavelength range the afterglow has faded below
 the brightness of the host. The dashed line shows the spectrum
 of SN 1998bw 24 days after the maximum (Patat \etal\ 2001), 
 redshifted to z=0.36 and
 diminished in brightness to 85\% (see \S 3.2.2), thus clearly 
 demonstrating that the supernova was too faint with respect to the host
 galaxy to be discovered spectroscopically. }
 \label{Ispec}
\end{figure}

During the maximum of the SN bump (Fig.~\ref{Allbandfit}),  the
observed brightnesses of the afterglow are 
$J$ = 22.8$\pm$0.1, $R$ = 23.4$\pm$0.1, and $V$ = 22.8$\pm$0.1.
This is about 3.6 mag, 3.3 mag, and 3.7 mag fainter than the total
light of the host galaxy (see below). Although not all of the host was 
covered by the slit, the host galaxy still dominated the collected light --
even for the spectrum taken on 12 December 2001, around the maximum
of the bump. Thus, the lack of any supernova signature in the spectrum
is no argument against the supernova interpretation of the bump
(Fig.~\ref{Ispec}), as pointed out by Garnavich \etal\ (2003).

The extinction-corrected (\S \ref{COBE}) absolute magnitudes  for
the supernova with a $k$-correction in the $VRJ$ bands of 0.85 mag 
(Leibundgut 1990), 0.55 mag and zero, respectively, are  
$M_{\rm J}$ = --19.18,
$M_{\rm R}$ =  --19.79, and $M_{\rm V}$ = --19.83.
Comparison with the absolute magnitudes of SN 1998bw suggest
that the supernova in GRB 011121 is slightly brighter in the $V$ band, 
but fainter at longer wavelengths.


\subsubsection{The GRB-supernova connection
\label{grbsncon}}

\begin{figure}[ht]
\vbox{\psfig{figure=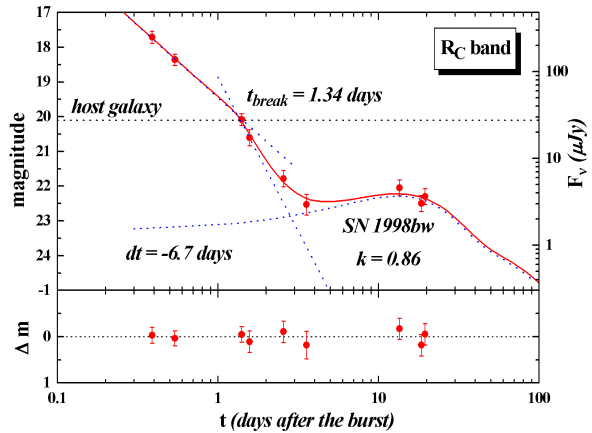,width=8.0cm,angle=0}}
 \caption[bspec]{The same as Fig.~\ref{Rcbandfit0} but with $dt$
  (eq.~\ref{dt}) as a further free parameter.}
\label{Rcbandfit}
\end{figure}

\begin{figure}[ht]
\vbox{\psfig{figure=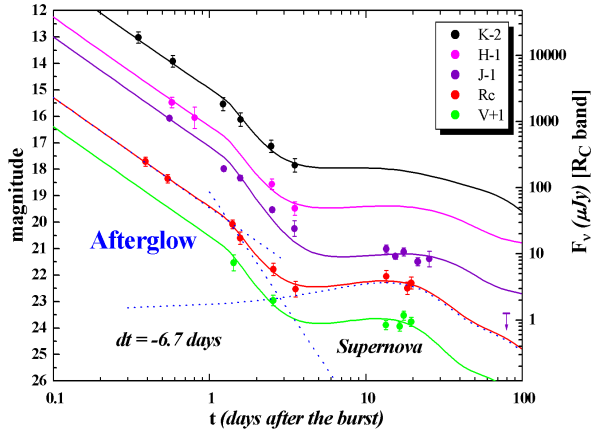,width=8.0cm,angle=0}}
\caption[bspec]{The same as Fig.~\ref{Allbandfit0} but for $dt=-6.7$
    days.}
\label{Allbandfit}
\end{figure}

While we use SN 1998bw as a template supernova, we added one degree of freedom
to the fitting procedure by allowing for a shift $dt$ of the observed SN
maximum  ($t_{\rm 011121}^{\rm max}$) with respect to the predicted one for a
redshifted SN 1998bw  ($t_{\rm 1998bw}^{\rm max}$), i.e.
\begin{equation}
           dt = t_{\rm 011121}^{\rm max} - t_{\rm 1998bw}^{\rm max}\,.
\label{dt}
\end{equation}
With this extra degree of freedom added, the  best fit parameters become:
$\alpha_1 = 1.67 \pm 0.58, \alpha_2 = 3.55 \pm 1.35, t_b = 1.34 \pm 0.60$
days, $k = 0.86 \pm 0.14$, and $dt = -6.7 \pm 5.2$ days with $\chi^2$/d.o.f. =
0.74. As before, the parameter $n$ was fixed at $n$ = 10 
(Figs.~\ref{Rcbandfit}, \ref{Allbandfit}).
The fact that $\alpha_2$ is less-well determined when we include the
data points  obtained after day 10 is not surprising. It is basically
due to the lack of data between day 4 and 10 combined with the early
and rapidly dominating SN component that makes the error  of the
deduced $\alpha_2$ relatively large. This is also the reason why in
Tab.~\ref{alphabeta}, where we discuss the appropriate afterglow
model, we used the fitting results obtained for those data when the SN
component is still negligible.


\begin{figure}[ht]
\vbox{\psfig{figure=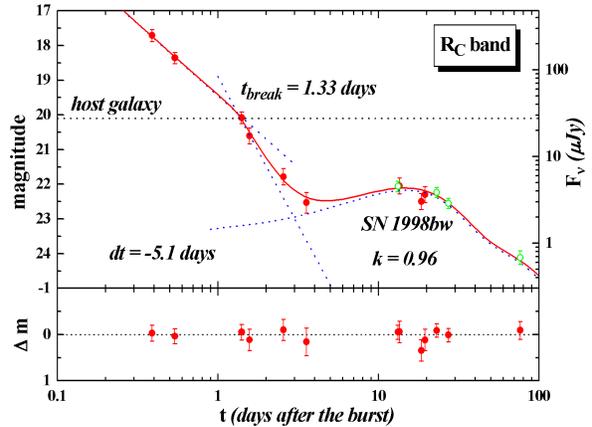,width=8.0cm,angle=0}}
 \caption[bspec]{The same as Fig.~\ref{Rcbandfit} but with the inclusion
  of published HST $R$-band data from Bloom et al. (2002).}
\label{HSTRcbandfit}
\end{figure}

The fit improves after inclusion of published late-time HST data 
(Bloom \etal\ 2002). With these additional data we find for the $R$-band 
light curve:
$\alpha_1 = 1.66 \pm 0.44, \alpha_2 = 3.43 \pm 0.68, t_b = 1.33 \pm 0.48$
days, $k = 0.96 \pm 0.06$, and $dt = -5.1 \pm 1.85$ days with $\chi^2$/d.o.f. =
0.60 (Figs.~\ref{HSTRcbandfit}). 
Finally, if we follow section~\ref{ohneSN} and include the transformed 
$H$-band data point and the early-time data from Garnavich et al (2003)
we get basically the same result with slightly reduced error bars
for $\alpha_1$ and $t_b$.

It was already previously noted (Bloom et al. 2002) that a negative time delay
$dt$ (Eq.~\ref{dt}) of order of a few days provides a
better fit to the data than setting $dt = 0$ days, i.e.  GRB and SN
start at the same time. Basically, two scenarios could explain such a
delay. First, this phenomenon could be intrinsic to the ejected
supernova shell itself in the sense that light curve shapes of type Ibc/II
supernovae are a function of the mass of the progenitor and other
details of the explosion (as already remarked by Bloom et
al. 2002). The more exciting alternative is that this delay could point
to a genuine time delay between the supernova explosion (formation of
a neutron star) and the GRB (interpreted as the subsequent formation
of a black hole; see Vietri \& Stella 2000). 
It is tempting to use
the negative delay found in our fitting procedure as an argument in support 
of the supranova model (Vietri \& Stella 2000).
However, it is worth to remember that the supranova model requires a
delay between the supernova and the GRB of many weeks to months  
(Vietri \& Stella 2000), not days. If this delay is shortened,
then the original goal of the model, namely to explain the
long-lived iron lines, is dismissed.
Much better data on both the GRB afterglow and the SN light curve are 
required to establish the reality of such an offset. GRB 011121 has not
been sampled enough to allow rigorous statements about the temporal 
relationship between SN and GRB, and the use of SN1998bw as a template
is also a rather unreliable assumption of our modeling, as pointed out
above, and further discussed below. 

\subsubsection{Alternative explanations for extra light}

We note that there are three observational details that do not fit the 
generally used template SN~1998bw:
\begin{enumerate}
\item Rapid intensity decay: The $J$-band flux on 9 February 2002 is far    
      below the prediction of SN 1998bw.  While there exist supernovae with
      rapidly decaying light curves, this faster decline applies only for
      times $t>80$ days.  These supernovae are thought to possibly produce
      less Ni, but more Ti. In the case of GRB 011121, the observed faster
      decay happened within less than 60 days after the maximum, and thus
      cannot be explained with a low Ni production.
\item Atypical color evolution: SN decay slower at longer wavelength,
      e.g., SN~1998bw or SN~2002ap.  The light curve of GRB 011121
      shows that the brightness decay in the  $J$ band is very rapid.
      Moreover, the HST data demonstrate that the decay in the $R$
      band is not only slower than that in the $J$ band, but also the
      $V$-band decay is slower than the $R$-band decay. Thus, in GRB
      011121 the color dependence is just inverted relative to a
      supernova: the longer the wavelength, the faster the decay.
\item Different spectral energy distribution: The excess light in
      GRB 011121 is substantially bluer than that of SN 1998bw,
      as already noted by Garnavich \etal\ (2003).
      This is most easily recognized in the different $k$ values
      for the different filters: while $k=0.86\pm0.14$ for the $R_c$ band
      (see previous section; excluding the HST data), we find 
      $k=1.10\pm0.50$ in the $V$ band
      and $k=0.40\pm0.03$ in the $J$ band. 
\end{enumerate}

We investigated the alternative to the supernova interpretation, namely a dust
echo due to scattered light (Esin \& Blandford 2000; Waxman \& Draine 2000;
Reichart 2001). In the absence of extinction within the host galaxy, as in the
case of GRB 011121, the dust echo is expected to be bluer than prescattered
light, just as observed.  This would suggest that a thermal dust echo is
clearly ruled out since it would peak in the near-infrared band 
(Reichart 2001). However, in the dust echo scenario there seems to
be a severe problem with the temporal evolution. Dust echoes due to
scattered light do not peak as sharply as thermal dust echoes or supernovae
(Reichart 2001). The time dependence comes from the angular dependence 
of the escape probability of the scattered
photons. Using the values of the differential escape probability as presented
by Esin \& Blandford (2000), we find that the inferred temporal decay  should
be much smaller than the observed decay rate after the bump maximum.  Other
alternatives are neutrons in the blast wave (Beloborodov \etal\ 2003), the
scenario of a refreshed relativistic shock in the GRB outflow caused by the
wind profile of the GRB progenitor (Ramirez-Ruiz \etal\ 2001), the shell
collision model (Kumar \& Piran 2000) or continuous injection (Bj\"ornsson
\etal\ 2002). However, these scenarios are not yet detailed enough to be
tested against observational data.

\subsection{The break in the light curve}

One might be concerned that the break in the R band light curve
happens at an intensity level which just corresponds to the
R band brightness of the host. However, this is chance coincidence,
and only applies to the $R$ band. There are three arguments
in favor of the break being unaffected by the host brightness:
(1) A subtraction of a constant flux could potentially produce a jump
  in the light curve, but no break with increasing deviation at later times.
(2) The R band brightness of the host is the brightness integrated
over the area of the host. In contrast, the additional flux which
the host would add to the PSF area of the afterglow is at least
a factor of 4 smaller. Thus, if the break would have been
``produced'' by a wrong host subtraction, it would occur at
a level nearly 2 magnitudes fainter. 
(3) The brightness of the host is different in different filter bands
  ($V-J = 1.3$ mag). This color is different than that of the afterglow
  ($V-J = 2.45$ mag; both values not corrected for extinction).
  Yet, the break is found at the same {\em time}
   in {\em all} filter bands, irrespective of the relative brightness
   of the host to the afterglow.

The break time $t_b$ we deduce has an 1$\sigma$\ error of  0.5 days so that one
might tentatively conclude that any break occurred between about 1 and 2 days
after the burst. Whereas a break at very early times ($t<1$ day) is excluded
by the data obtained by Price et al. (2002) and Garnavich et al. (2003), a
break at $t=2-3$ days  could be hidden by the bright SN component. The
following discussion relies on the deduced break time of $t_b \approx$ 1.3
days.

According to Livio \&
Waxman (2000), for a jetted explosion into  a wind-blown surroundings
with an $n(r) \sim r^{-2}$ gas density profile it is
\begin{equation}
\label{theta_jet}
       \Theta_{\rm jet}\approx 0.11 \
       \Big(\frac{E_{53}}{\dot{M}_{-5}/v_{\rm w,3}}\Big)^{-1/4} \
       \left(\frac{t_{b}}{1+z}\right)^{1/4}\,,
\end{equation}
where $\dot{M}_{-5}$ is the mass loss rate of the star in units of
$10^{-5}$ M$_\odot$ yr$^{-1}$, $v_{\rm w,3}$ is the wind velocity  in
units of $10^3$ km s$^{-1}$ and $E_{53}$ is the isotropic equivalent
energy of the fireball in units of $10^{53}$ erg.  We introduced
a factor $(1+z)$ in order to correct for the redshift. Using
$z$=0.36, $E_{53}$=0.27 and assuming $\dot{M}_{-5}/v_{\rm w,3}=1$  we obtain
$\Theta_{\rm jet} \sim 9\degr$. This corresponds to a beaming  factor
($b\approx2/\Theta_{\rm jet}^2$) of about 100.

Usually it is assumed that  $E_{53}$ can be approximated by the
isotropic equivalent energy release $E_\gamma$ in the gamma-ray band
during the burst phase.  According to Garnavich et al. (2003), for GRB
011121 it was  $E_\gamma=0.27 \times 10^{53}$ erg (see also \S 3.4). 
After correcting for the
beaming factor this gives an energy release of about $3 \times 10^{50}$ erg, 
in agreement with the typical energy releases of GRBs (Frail et al. 2001).

\begin{figure}[ht]
\vbox{\psfig{figure=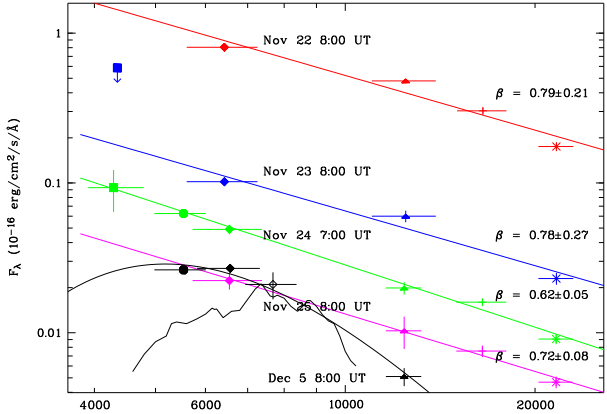,width=8.cm,angle=270,%
          bbllx=1.6cm,bblly=2.5cm,bburx=18.8cm,bbury=27.2cm,clip=}}\par
\caption[broad]{Extinction-corrected, broad-band spectral energy 
   distribution of  GRB
   011121 on four consecutive days, starting about 14 hrs after the GRB,
   and on 5 December 2001 near the maximum of the supernova-bump.
   The best-fit power law slopes $\beta$ are given at the right side
   except for December 5.
   The maximum deviation of the data to the power law is
   $\sim$0.15 mag, justifying the systematic error added in the light
   curve fitting procedure.
   It can be seen that the spectral slope does not change over the first
   four days, supporting the jet nature of the break seen in the light curve. 
   The December 5 (filled circles) spectrum is clearly not of power law type. 
   The smooth
   solid curve is a blackbody with a temperature of 6300 K, while the
   thin solid line is the spectrum of SN 1998bw, redshifted to $z$=0.36.
   Note that on  Nov. 24, approximately 1 day after the break,
   the value of $\beta$ is well-defined and within the error bars well
   consistent with the predictions of the jet-wind model 
   (Table~\ref{alphabeta}; case 2). 
\label{broadband}}
\end{figure}

Given the many multi-filter observations during the first 4 days
(Tab. \ref{log}), we have constructed broad-band spectra from the  optical to
the near-infrared region. These are shown in Fig. \ref{broadband} with a
separation of about 24 hrs between each other. The spectral energy
distributions of the first four days are consistent with a 
$\beta = 0.70$ power-law. 
This suggests that the break in the light curve was achromatic, in agreement 
with the predictions of the jet model.
We note in passing that we do not see $\beta$ to evolve towards
an asymptotic value of 1.1 as has been argued by Dado \etal\ (2002b)
and assumed for the cannonball interpretation of the GRB 011121
afterglow light curve.


\subsection{The host galaxy \label{hostgalaxy} }

\subsubsection{Morphology}

Images in $R, V$, and $J$ show that the host galaxy starts affecting
the brightness estimate of the afterglow already after November 24,
2001.  We have therefore taken the $J$-band image from February 9 to
model the shape and intensity distribution of the host galaxy in order
to subtract the host flux from the earlier images. The February 9
exposure consists of two sub-exposures, each with 30 min exposure
time, and  taken at a mean seeing of 0\farcs45 and 0\farcs55,
respectively.  We used the 0\farcs45 part and modelled the host galaxy
using a bulge component
\begin{equation}
          I_{\rm bulge}(R) =  I_{\rm bulge,0} \ \exp \Big[-7.67\:
\Big(\frac{R}{R_e}\Big)^{1/4}\Big]
\end{equation}
and a disk component
\begin{equation}
   I_{\rm disk}(R) = I_{\rm disk,0} \ \exp \Big(-1.68\:
   \frac{R}{R_e}\Big)\,.
\end{equation}
For each of these two components, the four parameters $I_{\rm
bulge,0}$, ($I_{\rm disk,0}$),  effective radius $R_e$, position angle
(PA) and eccentricity  ($e = \sqrt{1 - (b/a)^2}$) are derived using the 
IRAF/SPP package GIM2D (Simard 1998).  
The resulting fit parameters are:
\begin{itemize}
\item Bulge component: $I_{\rm bulge,0}$ =  14.78 mag/arcsec$^2$,
      $R_e$ = 5.0 pixel,  $e$ = 0.19, PA = --41\grad
\item Disk component:   $I_{\rm disk,0}$ = 22.18 mag/arcsec$^2$, $R_e$
      = 19.5 pixel, $e$ = 0.19, PA = --46\grad\,.
\end{itemize}
The eccentricity derived from the galaxy profile in the $J$ band
corresponds to an inclination of $\sim$10 degrees.

\begin{figure}[ht]
\vbox{\psfig{figure=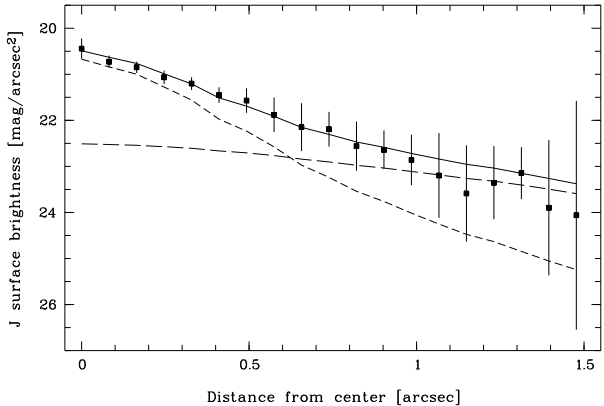,width=8.cm,angle=270,%
          bbllx=1.6cm,bblly=2.5cm,bburx=18.8cm,bbury=27.2cm,clip=}}\par
\caption[lc]{Surface brightness profile of the host galaxy of  GRB
   011121 in the $J$ band, measured on the 2002 February 9 image  (first
   part) at 0\farcs45 seeing with a plate scale of 0\farcs147/pixel
   (two data points per pixel).  
   The abscissa is the distance from the
   center measured as $\sqrt{a^2+b^2}$, with $a$ and $b$ being the major
   and minor axes of the galaxy.  The dashed line denotes the bulge
   component, the long-dashed line the disk component, and the full line
   the sum of both components.
\label{galprof}}
\end{figure}

The fit to the radial profile is given in Fig. \ref{galprof}.  To
subtract the host galaxy emission from the images taken in
November/December 2001, this model of the host galaxy was convolved
with  the corresponding seeing of each image, and then subtracted.
For filters other than $J$, the total brightness of the model was
modified to achieve good subtraction in the corresponding band,
i.e. the normalization was varied to minimize the residuals.
Aperture photometry on the subtracted images was then applied to
obtain the magnitudes of the afterglow (see Tab. \ref{log} and Fig.
\ref{Allbandfit0}).

Using the model parameters given above, and integrating over 
the galaxy, we obtain the following apparent magnitudes of the host galaxy: 
$J$ = 19.1$\pm$0.1, $R$ = 20.1$\pm$0.1, and $V$ = 20.4$\pm$0.1.


The decomposition of the surface brightness profile of the galaxy into
a disk and a bulge component shows that the bulge dominates within the
inner 0\farcs5 only. The bulge contributes about 22\% to the total
light.  This can be interpreted as  a Sbc morphological type.  The
observed brightnesses of the host galaxy in the $V$, $R$ and $J$ band
correspond, after correction for foreground absorption and applying
the $k$-correction,  to absolute magnitudes of --22.58$\pm$0.53, 
--22.24$\pm$0.24,  and --22.09$\pm$0.25, respectively. This is at the very
bright end of the range populated by spirals (e.g., Binggeli \etal\
1988); the host galaxy of GRB 011121 is thus among the brightest 5\%
Sbc galaxies.

The gamma-ray burst happened at an offset of 0\farcs9 from the center of the
host galaxy (as measured on a VLT image taken at 0\farcs6 seeing),
corresponding to a physical distance of 9 kpc (see also Ryder \etal\ 2001). 
The above derived 
disk radius is $R_e$ = 29 kpc.
This offset location of the gamma-ray burst with its supernova is
consistent with the general supernova picture, as well as with
offsets found for earlier GRBs (Bloom \etal\ 2002).

\subsubsection{The star-formation rate}

Using the extinction corrected line fluxes of [O~II] and
H$\alpha$ (Tab. \ref{lflux}) and the relations given by Kennicutt (1998),
i.e. 
SFR (\mdot/yr) = 1.4$\times$10$^{-41}$ $\times$ L(OII) and  
SFR (\mdot/yr) = 7.9$\times$10$^{-42}$ $\times$ L(H$\alpha$),
we deduce a star
formation rate of 1.2 \msun/yr and 0.72/0.61  \msun/yr, respectively.
In contrast, Subrahmanyan \etal\ (2001) derive a star-formation
rate of 13--70  \msun/yr from their radio detection of the host
galaxy at 0.05 mJy and assuming a spectral slope of --0.5 
between 1.4 and 8.5 GHz.
The rate derived here is at the low end of the range found previously 
in GRB host
galaxies. However, also for these earlier cases different results
are obtained for different methods/lines, in particular also
a large difference between  radio- vs. optically-based data 
(Vreeswijk \etal\ 2001).

We have no explanation for the fact that the host galaxy appears
to be among the brightest 5\% Sbc galaxies, and yet the derived
star-formation rate is rather low.
Also, 
the lack of clear evidence for dust in the GRB host along
the line of sight is somewhat surprising 
in the supernova picture, though there is also no strong intrinsic 
extinction in most previous GRB hosts.

\section{Concluding remarks}

While our $\alpha$ and $\beta$ values are consistent within the errors 
with those reported by Garnavich \etal\ (2003) and Price \etal\ (2002),
the change in our values due to the discovery of the break 
and the evolution during the four days after the GRB leads
to a preference of the jet-wind model. 
The observed break and its interpretation is in contradiction with
the interpretation of the radio data by Price \etal\ (2002)
suggesting a break at times later than 8 days after the GRB.
We cannot offer a solution to this problem, as the break is
observationally evident from a dozen of data points spread over several
filter bands.

Our observations reveal a light curve  break at early
times and the appearance of extra light at late times. 
The former is believed to be evidence for a collimated outflow (a
fireball jet), whereas the latter could be due to supernova light.
Finally, the $\alpha-\beta$-relations favor a wind model 
(Table~\ref{alphabeta}). The afterglow of GRB 011121 thus provides another
case in support of the current standard paradigm of long-duration
gamma-ray bursts: the collapsar model (Woosley 1993, 
Hartmann \& Woosley 1995) in which the 
birth of a black hole inside a rapidly rotating massive star is
announced via jet formation, breakout, and propagation into a stellar
environment that was shaped by the strong winds of the pre-collapse
star. In fact, GRB 011121 is the first case, where all three
signatures (jet break, wind density profile and supernova bump)
have been found, while in earlier cases only two of these three
ingredients could be found (e.g., Jaunsen \etal\ 2001).
However, even with GRB 011121 this line of reasoning remains
circumstantial, and many more afterglows have to be observed, and 
better sampled than is presently the case, to seriously establish 
the GRB-collapsar picture.

\section{Acknowledgments}

We are highly indebted to the ESO staff, in particular  N. Ageorges,
S. Bagnulo, H. B\"ohnhardt, V. Doublier, O.R. Hainaut, S. Hubrig,
A.O. Jaunsen, E. Jehin, R. Johnson, A. Kaufer, M. K\"urster,
E. Mason, E. Pompei, Th. Szeifert for prompt execution of the
observing requests and all the additional effort related to that.
We thank the anonymous referee for the extensive comments on the
original version of this paper.
J. Gorosabel acknowledges the receipt of a Marie Curie Research Grant from  
the European Commission, and J.M. Castro Cer\'on the receipt of a FPI 
doctoral fellowship from Spain's Ministerio de Ciencia y Tecnolog\'{\i}a.
This work was supported by the Danish Natural Science Research Council (SNF),
by ``IUAP P5/36''
Interuniversity Attraction Poles Programme of the Belgian Federal Office for
Scientific, Technical and Cultural Affairs, and the Belgian Fund for
Scientific Research (FWO).
This research has made use of the USNOFS Image and Catalogue Archive
 operated by the United States Naval Observatory, Flagstaff Station
 ({\tt http://www.nofs.navy.mil/data/fchpix/}).


\end{document}